\documentclass[aps,twocolumn,superscriptaddress,showpacs]{revtex4} 

\newcommand{\beq}{\begin{equation}}
\newcommand{\eeq}{\end{equation}}
\newcommand{\test}{\mbox{$\begin{array}{c}\stackrel{\stackrel{\textstyle \mathcal{H}_1}{\textstyle >}}
{\stackrel{\textstyle <}{\textstyle \mathcal{H}_0}} \end{array}$}}
\newcommand{\hg}{  }
 \usepackage{epsfig}
\usepackage{natbib}
\usepackage{amsmath}
\usepackage{times}
\usepackage{psfrag}
\usepackage{epstopdf}

\usepackage{graphicx}
\usepackage{subfigure}
\usepackage{float}

\usepackage{hyperref}

\begin{document}


\title{ Interplay between Detection Strategies and Stochastic Resonance Properties}

\author{Paolo Addesso}
\address{Dept. of Information Engineering, Electrical Engineering and Applied Mathematics
(DIEM), University of Salerno, Via Giovanni Paolo II, 132, I-84084, Fisciano, Italy.
} 
\author{Vincenzo Pierro}
\address{Dept. of Engineering, University of Sannio, Corso Garibaldi, 107, I-82100 Benevento, Italy}
\author{Giovanni Filatrella $^1$}
\address{Dept. of Sciences and Technologies, 
University of Sannio, Via Port'Arsa, 11, I-82100 Benevento, Italy} 
\begin{abstract}
We discuss how to exploit stochastic resonance with the methods of statistical theory of decisions. 
To do so, we evaluate two detection strategies: escape time analysis and strobing.
For a standard quartic bistable system with a periodic drive and disturbed by noise,
we show that the detection strategies and the physics of the double well are connected, inasmuch as one (the strobing strategy) is based on synchronization, while the other (escape time analysis) is determined by the possibility to accumulate energy in the oscillations.
The analysis of the escape times best performs at the frequency of the geometric resonance, while strobing shows a peak of the performances at a special  noise level predicted by the stochastic resonance theory. 
We surmise that the detection properties of the quartic potential are generic for overdamped and underdamped systems, in that the physical nature of resonance decides the competition (in terms of performances) between different detection strategies.\\
\end{abstract}
\pacs{05.40.-a,02.50.-r,84.40.Ua,07.50.Qx
{\it  To appear on Commun. Nonlinear Sci. Numer. Simulat. (2015)}
}

\maketitle


\section{Introduction}

It is well known that Stochastic Resonance (SR) can be exploited, under suitable circumstances,  to improve detection \cite{Benzi81,McNamara89, Gammaitoni98,Galdi98}. 
However, the knowledge that the system response could possibly be enhanced at special noise level is not sufficient to identify the best method to be employed for signal detection. 
In fact, several indexes have been proposed to quantify the response of a system when SR occurs: e.g.,  Fourier spectrum analysis \cite{Benzi81}, \cite{McNamara89}, synchronization measures through strobed data \cite{Galdi98}, amplification of a reaction coordinate \cite{Gammaitoni89}, escape times processing \cite{Fauve83,Bulsara03,Wellens04}. 
Therefore, it would be interesting to know in advance which method will actually work better.
Put it in another way, it is interesting to know if a link exists between the  physics characteristics of the system and the performance behavior  of feasible detection strategies. 

Let us recall some basic concepts of SR from the point of view of statistical decision theory.
To be specific,  we consider a signal $S(t)$ that is a mixture of a periodic drive of amplitude $\alpha$, (angular) frequency $\omega$, initial  (unknown) phase $\varphi_0$, and a random (uncorrelated Gaussian) perturbations $\xi$:
\beq
S(t) = \alpha \sin \left( \omega t +\varphi_0\right) + \xi (t)
\label{signal}
\eeq
\noindent The conventional wisdom is that Stochastic Resonance (SR) occurs when the response of a nonlinear system to the signal $S(t)$ can be enhanced at a {\it special} noise level. 
In the standard analysis of SR \cite{McNamara89,Gammaitoni98,Wellens04}, the response is evidenced by the behavior of a component of the outgoing Fourier spectrum, that is maximized at a certain noise intensity. 
The upsurge of the Fourier component makes SR appealing for signal detection, inasmuch as it is conceivable to exploit the Fourier analysis to reveal the presence of {\hg an injected deterministic } signal \cite{Inchiosa96}. In this scheme, one hopes that  a particular combination of noise and periodic {\hg drive} makes it easier to detect the signal, for instance because a threshold is only passed with the help of the noise \cite{Ando01} (although it has been proposed to exploit stochastic resonance also for  suprathreshold {\hg deterministic signals} \cite{Apostolico97}).
However, the very concept that noise could be beneficial is counterintuitive, if not controversial \cite{Simonotto97,Mcdonnell09}, in that it sounds against good sense that \emph{more} disturbances can, in the end,  improve the detection of a signal. 
This Gordian knot  has been cut by noticing that SR only helps detection in \emph{suboptimal} systems \cite{Kay00, Rousseau05},  and suboptimal threshold choices \cite{Ward02}. 
In less general terms, let us suppose we want to decide about the presence of a periodic drive; then decision theory proves that the Neyman-Pearson scheme is optimized by the Likelihood Ratio Test (LRT). 
To decide about the presence of a periodic forcing corrupted by white Gaussian noise LRT amounts to the scalar product of the signal $S(t)$ and the mask, the periodic component $\sin \left( \omega t \right)$. 
This is the optimal detection strategy -- the matched filter applied to the input  signal $S(t)$ \cite{Helstrom68} -- that cannot be improved adding noise. 
In practice, one is often frustrated in the application of LRT optimal technique and employs suboptimal strategies where SR based signal detection enhancement can genuinely occur.
As preeminent examples where the optimal strategy is not feasible, we can mention the cases where the full trajectory $S(t)$ is difficult to retrieve (as in very sensitive Fabry-Perot pendulums for gravitational waves detection \cite{Addesso13,Addesso14}), or it is just not available for measurements (as in Josephson junctions, for the quantum mechanical nature of the dynamical variable \cite{Hibbs95,Addesso12,Valenti14}). In some other cases \cite{Krolak98} the recorded signal is far too long to be analyzed with optimal methods, and the LRT is {\it de facto} not applicable.
In still other cases a nonlinearity transformation in the system allows for the occurrence of typical SR pattern \cite{Ward02}.
To visualize the difficulty, we can imagine to tackle the original problem where SR arose: the study of climate changes, 
based on geological evidence of the alternate of dry and cold periods \cite{Benzi82}. 
Can we access the instantaneous temperature of the Earth, that is the signal $S(t)$? Unfortunately, we can just estimate the passage from an ice age to a dry age, i.e. the escape time from stable climate configurations. 
Even if better techniques were available to estimate the yearly Earth temperature, say from the maximum extension of continental ice sheets, this amounts to measure each year the temperature in the coldest day. 
Put another way, one could only observe the so-called strobed dynamics obtained illuminating the system at some time intervals. 
The two sampled dynamics we have just mentioned -- Escape Times (ET) from dry to cold periods and Strobed Dynamics (SD) at prescribed time intervals -- are suboptimal, in that the optimal LRT strategy, equivalent to the matched filter, requires to exploit the whole trajectory (i.e., the full information content, or the instantaneous temperature)  of the input signal. 
Only for the suboptimal strategies, based on the reduced data (e.g.  the escape times or the strobed dynamics), SR can occur.
Indeed it has been shown that: i) noise can be used to enhance  signal detection through the analysis of ETs in the first order standard bistable potential \cite{Wellens04} and in a second order washboard potential \cite{Filatrella10}
ii) with the appropriated choice of noise intensity SD exhibits good detection performances, $\approx 3 dB$ below the optimum, when strobing occurs at the forcing period $2\pi/\omega$  \cite{Galdi98}.

\begin{figure}
\centerline{\includegraphics [scale=0.5]{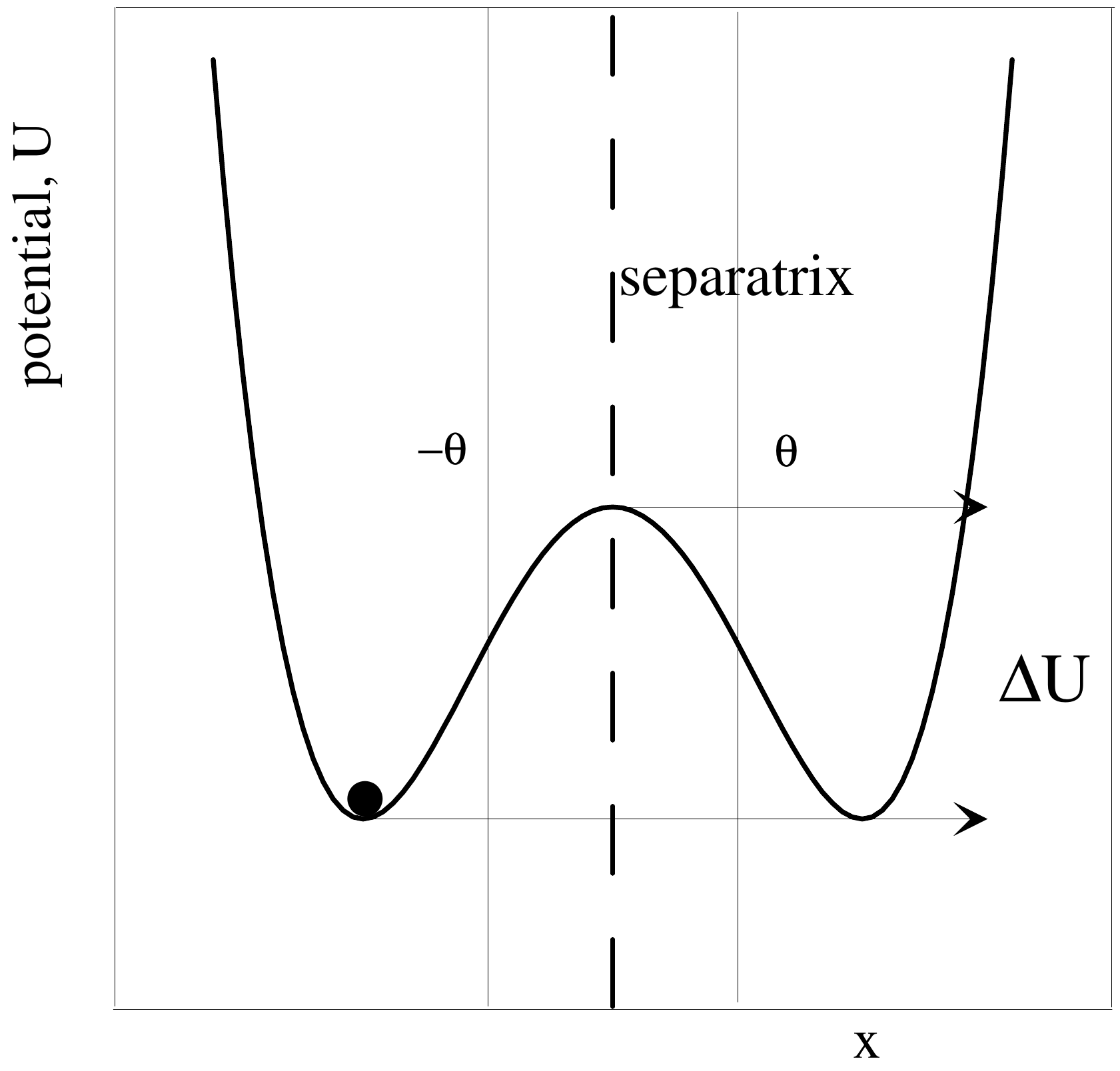}}
  \caption{Sketch of the escape process. The separatrix for the potential Eq.(\ref{eq:bistapot}) is at $x=0$. The limits $\theta$ and $-\theta$ are discussed in the Appendix.}  
\label{fig:potential}
\end{figure}

\begin{figure}
\centerline{\includegraphics [keepaspectratio,width=9cm]{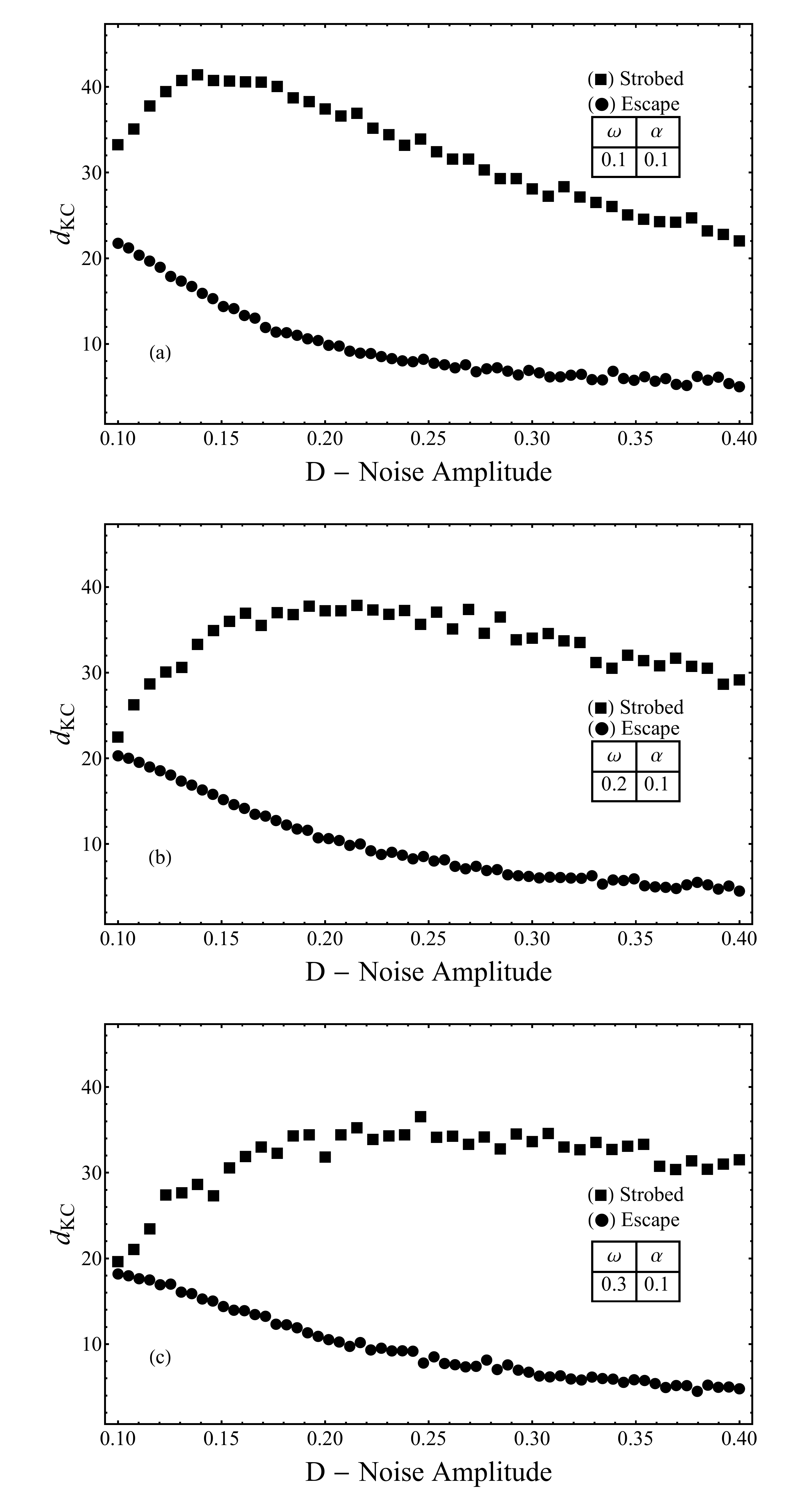}}
  \caption{Performances of the detection for the overdamped system as measured by the KC index $d_{KC}$ for the ET and SD strategies. 
 The figure demonstrates a decrease of the performances of the escape, without resonance or peak, and independent of the frequency of the {\hg periodic drive} $\omega$, in the considered noise interval ($0.1 \le D \le 0.4$).
  The peak of the strobed technique decreases when the frequency increases, see also Fig. \ref{indicaSRdiversi}a. 
 }
\label{defVSstrobALL}
\end{figure}

Thus the alternatives to the matched filter are practically attractive, and could possibly exhibit a {\it bona fide} enhancement when noise increases. 
 If one backs down the optimal matched filter, physical intuition suggests to seek for best performances in the parameter region where SR occurs.
Put it another way, if  an opportunity to improve the analysis by adding noise is to exist, one guesses that the resonant condition of SR is the first place where to look for such a chance.

Still, an open problem can be posed: which technique, among the many suboptimal ones, best performs in a specific physical system? 
The objective of the present work is to systematically characterize the two detection performances of  two reduced data -- the escape times and the strobed dynamics -- for the overdamped and underdamped prototypal system of a quartic potential. 
The analysis of the performances demonstrates that the two techniques display different properties, and that the best performing strategy depends upon the underlying dynamical nature of the resonance.
We remark that we have selected two particular strategies which are suitable when the full trajectory is not easily available. 
Other suboptimalities can be devised, for example threshold detectors, similar to a neuron,  that are most suited in biological applications \cite{Ward02,Fiasconaro08}.

The paper is organized as follows: in Sect. \ref{sec:model} we outline the model equations. In Sect. \ref{sec:detection} we establish the suboptimal detection strategies for ET and SD, that are applied and evaluated for the prototypal quartic potential in Sect. \ref{sec:performances}. Sect. \ref{sec:conclusions} concludes with the physical consequences of the above analysis.

\begin{figure}
\centerline{(a)}
\centerline{\includegraphics [keepaspectratio,width=8cm]{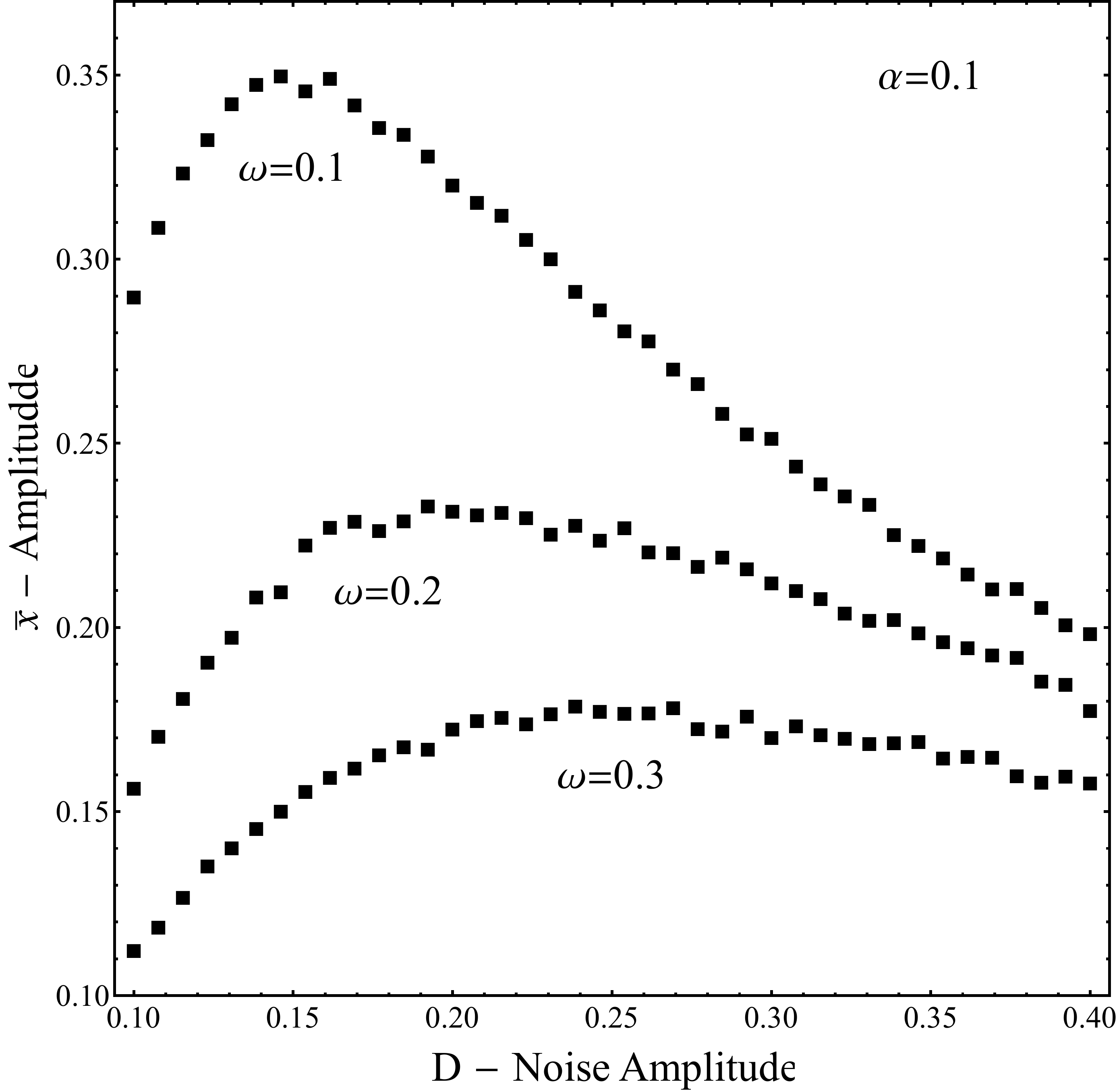}}
\centerline{(b)}
\centerline{\includegraphics [keepaspectratio,width=8cm]{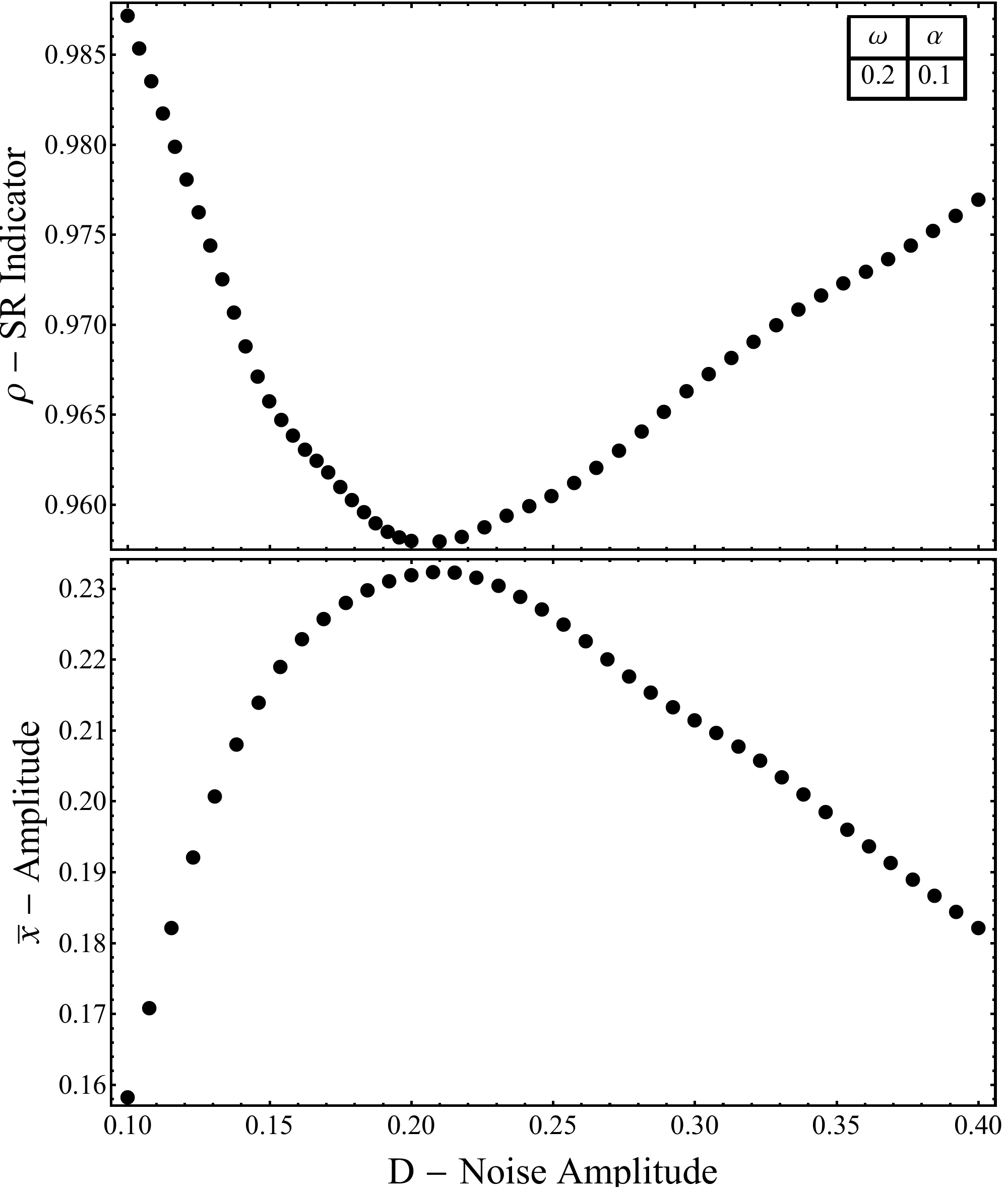}}
  \caption{Comparison of different indicators of SR, obtained by simulations of Eq.(\ref{eq:bistafirst}). (a) The average amplitude of the displacement, Eq.(\ref{displacement}), that decreases with the external frequency. 
  (b) Distortion of the ETs distribution, Eq.(\ref{rho}), compared to the displacement for the middle frequency, $\omega = 0.2$ (the data have ben smoothed to evidenced the peak maximum and the minimum).
This is the standard SR, as the noise at which the peak occurs increases with the external frequency. 
Parameters are the same as in Fig. \ref{defVSstrobALL}.
}
\label{indicaSRdiversi}
\end{figure}

\begin{figure*}
\hspace{2.7cm}(a) \hspace{6.7cm}(b)\\
\centerline{\includegraphics [keepaspectratio,width=14cm]{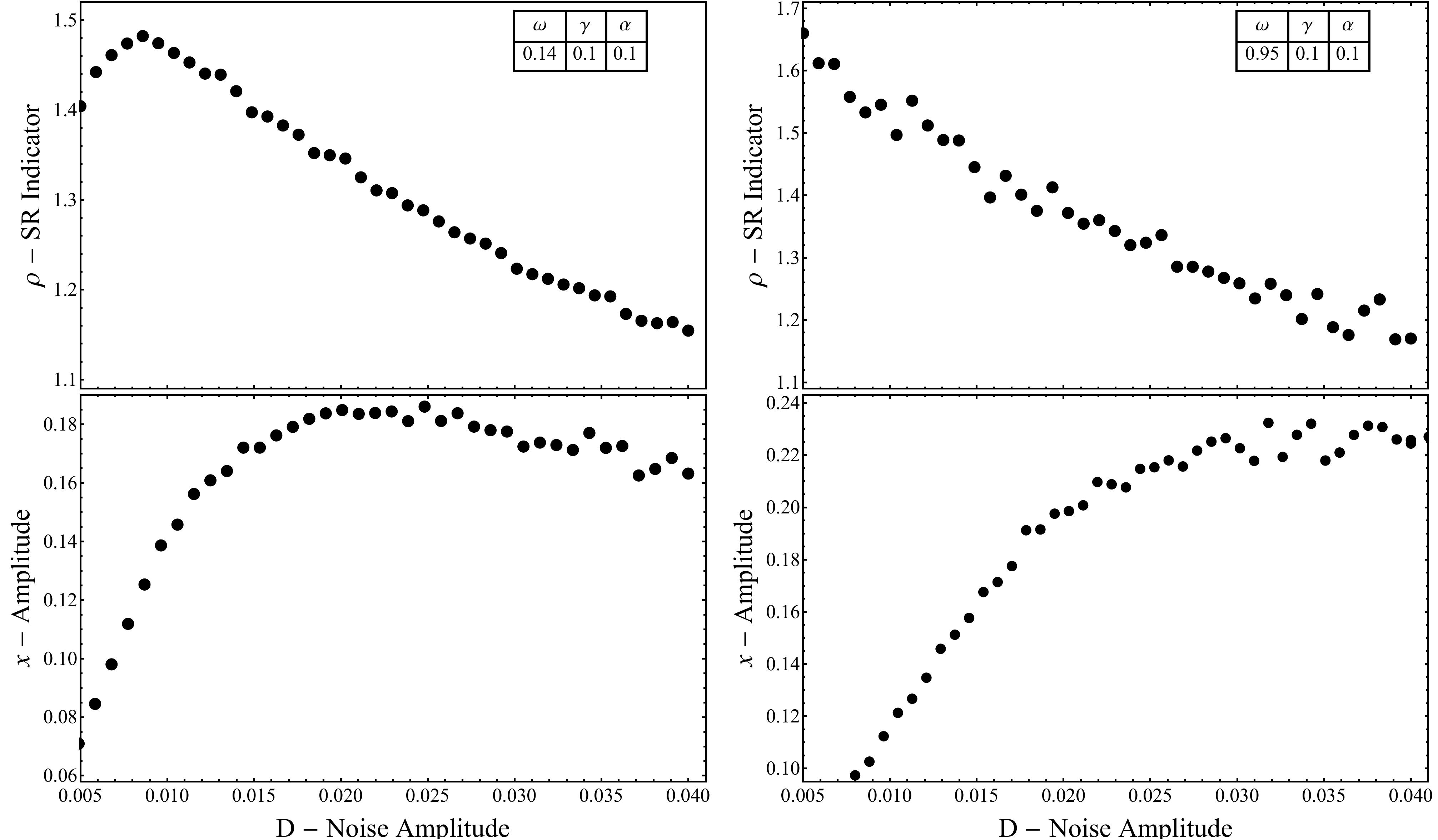}}
\caption{
Comparison of different indicators of SR, obtained by simulations of Eq.(\ref{eq:bistasecond}): Distortion of the ETs distribution, Eq.(\ref{rho}), 
and average amplitude of the displacement, Eq.(\ref{displacement}). 
(a) Low frequency behavior, $\omega \ll 1$. The indexes exhibit a maximum in correspondence of the best performances, see next Fig. \ref{icastica}.
(b) Geometric resonance frequency behavior, $\omega \simeq 1$. The indexes exhibit a monotonic behavior typical of the regions without stochastic resonance.  
}
\label{icasticaindex}
\end{figure*}

\section{ Models for Bistable Systems}
\label{sec:model}

\subsection{First Order Model }
\label{subsec:firstorder}
Let us consider the signal  (\ref{signal}) applied to a prototypal quartic bistable potential; in normalized units the system is governed by the following stochastic differential equation \cite{Gammaitoni98}:

\beq
\frac{dx}{dt} - x  + x^3 =  \alpha \sin(\omega t+\varphi_0)+\xi(t) = S(t).
\label{eq:bistafirst}
\eeq

\noindent
Eq. (\ref{eq:bistafirst}) is called overdamped because is the high friction limit of a nonlinear oscillator driven by a mixture of {\hg deterministic signal} and noise, as will be discussed in the next Subsection. 
In Eq. (\ref{eq:bistafirst}) a random terms appears, $\xi(t)$, to model an additive noise with autocorrelation function of intensity $D$, viz $\langle\xi(t)\xi(t')\rangle=2D\delta(t-t')$, corrupting the external sinusoidal drive.
 Thermal noise has been supposed uncorrelated with external noise and included in the overall noise $D$. For practical applications the physical nature of the noise sources (either intrinsic thermal noise or external noise) does affects the mathematical treatment  of the detection strategies (if the sources are uncorrelated).  
We focus on the simple case of additive noise (that is a paradigm in signal processing); however we expect that the results can be indicative of the behavior also for other noise sources and systems, as non-Gaussian \cite{Fuentes2001,Lacognata10}, colored noise \cite{Gammaitoni89b,Gammaitoni93} in inertial and overdamped systems \cite{Hanggi1993}, combinations of time delay and non-Gaussian noise \cite{Masoller01,Masoller02,Wu07} 
{\hg and with different mechanisms \cite{Borromeo04,Borromeo07}. }

The SR frequency is (approximately) the following:
\beq
\langle\tau \rangle _0 \simeq \frac{\pi}{\omega_{SR}}.
\label{eq:SR}
\eeq
\noindent Here $\langle \tau \rangle$ is the average of the ETs when the noise intensity is $D$ (angular brackets denote time average, the subscript $0$ indicates the absence of the {\hg external drive}).

The quartic bistable potential  of Fig. \ref{fig:potential} is associated with Eq. (\ref{eq:bistafirst}):

\beq
U(x)=\frac{x^4}{4}-\frac{x^2}{2}.
\label{eq:bistapot}
\eeq

If the periodic {\hg deterministic signal} is absent ($\alpha=0$), for low noise  ($D<<\Delta U$) the escapes occur at a rate \cite{Risken} 
\beq
\langle \tau \rangle ^{-1} = \tau_K^{-1} \exp\left(-\frac{\Delta U}{D}\right)
\label{kramer}
\eeq
where $\tau_K$ is the Kramers prefactor \cite{Risken}, {\hg that also gives the order of magnitude of the relaxation time of local equilibrium \cite{Lanzara97}.} The normalized barrier height is $\Delta U = 1/4$ for the normalized potential (\ref{eq:bistapot}). Thus, when the {\hg external drive} is absent, Eq.(\ref{kramer}) represents the expected average.

The average escape time $\langle\tau\rangle$ can be directly measured in experiments\cite{Benzi81,Gammaitoni89}, and in many cases it is the only {\em physical} quantity promptly available \cite{Hibbs95,Addesso12,Addesso13}. 
In fact in some physical systems as Fabry-Perot pendulums or Josephson junctions  \cite{Addesso13,Addesso12}  it is difficult (or impossible) to follow the entire trajectory of the system, while it is possible to detect the passage across a separatrix (and therefore the sequence is more rigorously defined as a passage time \cite{Risken}).
In the presence of the drive, the average passage time is altered by the sinusoidal term. 
Loosely speaking, one could therefore distinguish if the {\hg deterministic signal} is present ($\alpha \neq 0$) or absent ($\alpha=0$) by means of the escape rate, as the two cases correspond to $\langle \tau \rangle \neq \langle \tau \rangle_0$ or $\langle \tau \rangle = \langle \tau \rangle_0$, respectively (a rigorous formulation of the problem will be discussed in Sect. III). SR leads to a more effective signal detection if an increase of the noise $D$ facilitates the distinction between the two responses (with and without {\hg external drive}) . 

\subsection{Second Order Model}
\label{subsec:secondorder}
The bistable model Eq.(\ref{eq:bistafirst}) can be straightforwardly generalized introducing inertia and damping ($\gamma$), that lead to the following model:
\beq
\frac{d^2x}{dt^2}+\gamma \frac{dx}{dt} - x + x^3  =  \alpha \sin(\omega t+\varphi_0)+\xi(t) = S(t).
\label{eq:bistasecond}
\eeq

Equation (\ref{eq:bistasecond}) is a second order stochastic differential equation of a nonlinear bistable oscillator in a potential given by Eq.(\ref{eq:bistapot}). 
For $\gamma < 1$ this amounts to an underdamped or inertial system.
The normalizations of Eq.(\ref{eq:bistasecond}) are not the same as per Eq.(\ref{eq:bistafirst}), thus the two systems cannot be directly compared; however, we will focus on some features that are not affected by the different normalizations.

The analytical treatment of the stochastic differential Eq.(\ref{eq:bistasecond}) is much more difficult than the overdamped case, Eq.(\ref{eq:bistafirst}), and a general solution valid for all dissipation values has not been found, yet \cite{Berglund05}. 
Even in the absence of the {\hg periodic drive}, the unperturbed escape rate is only obtained with some approximations, therefore the theoretical understanding of second order SR is less complete than in the case of first order system. 
At a simple level, however, the interpretation of SR for signal detection is the same: subject to a sinusoidal excitation the escape rate  moves away from the unperturbed value, and such displacement is enhanced for special values of the random intensity \cite{Gammaitoni89b}.
 
\section{Detection Strategies}
\label{sec:detection}
To fully define the detection strategies, it is usual to formalize the problem as a binary hypothesis test:
\[
\begin{array}[c] {cl}
\mathcal{H}_0: & \text{sinusoidal drive is absent}\\
\mathcal{H}_1: & \text{sinusoidal drive is present}
\end{array}
\]
For this decision problem two different error probabilities arise:
\begin{itemize}
\item the \emph{false alarm probability} $P_f$, also called Type I error probability, i.e. the probability to decide for the hypothesis $\mathcal{H}_1$ when $\mathcal{H}_0$ is true;
\item the \emph{miss probability} $P_m$, also called Type II error probability, i.e. the probability to decide for the hypothesis $\mathcal{H}_0$ when $\mathcal{H}_1$ is true.
\end{itemize}

We first consider the case of a bistable system where the normalized potential $U(x)$  and the normalized noise standard deviation $D$ are known and do not depend on the particular hypothesis $\mathcal{H}_0$ or $\mathcal{H}_1$ in force. 
We also assume that the signal parameters (i.e. $\alpha$, $\omega$, and $\varphi_0$) are known under $\mathcal{H}_1$ hypothesis.
In this setup the Neyman-Pearson lemma \cite{Helstrom68} identifies the LRT as the optimal detection strategy, for it minimizes, among all possible tests, the miss probability $P_m$ at a fixed false alarm level $P_f$.
Thus, for a fixed mean observation time $\langle T \rangle$, if we collect $M$ observations $\underline{y} = \left\{y_k\right\}_{k = 0}^{M-1}$, supposed to be independent and identically distributed, the test statistics can be written as:
\beq
\displaystyle{\prod_{k = 0}^{M-1}\frac{f_Y(y_k|\mathcal{H}_1)}{f_Y(y_k|\mathcal{H}_0)}} 
\test
\zeta^{\prime},
\label{eq:likeT}
\eeq
where $f_Y(\cdot|\mathcal{H}_{0,1})$ are the Probability Density Functions (henceforth PDF) of the observation $y$ under the hypothesis $\mathcal{H}_{0,1}$, while $\zeta^{\prime}$ is a suitable threshold that returns a fixed false alarm level. 
To simplify the computation of the statistics (\ref{eq:likeT}), it is useful to compare the normalized natural logarithm of the likelihood ratio (henceforth LLR) with a threshold $\zeta = \log(\zeta^{\prime})$:
\beq
\Lambda(\underline{y}) = \displaystyle{\sum_{k = 0}^{M-1}\log \left[ \frac{f_Y(y_k|\mathcal{H}_1)}{f_Y(y_k|\mathcal{H}_0)} \right]}
\test
\zeta.
\label{eq:loglikeT}
\eeq
The advantage of Eq. (\ref{eq:loglikeT}) is that the statistic $\Lambda(\underline{\tau})$ can be computed as the sum of the random samples 
$\underline{\mathcal{L}} = \left\{\mathcal{L}_k\right\}_{k = 0}^{M-1}$, that are obtained from the observation via the \emph{optimal} (in the Neyman-Pearson sense), transformation
\beq
\mathcal{L}_k = \log \left[ \frac{f_Y(y_k|\mathcal{H}_1)}{f_Y(y_k|\mathcal{H}_0)} \right]
\label{eq:locloglik}
\eeq
that contains the information of both PDFs, $f_Y(\cdot|\mathcal{H}_1)$ and $f_Y(\cdot|\mathcal{H}_0)$.

To further simplify the analysis we employ the  Kumar-Carroll (KC) index $d_{KC}$ \cite{KC}: 
\beq
d_{KC}(\Lambda)= \frac{\mid \langle\Lambda\rangle_1-\langle\Lambda\rangle_0\mid}{
\sqrt{ \frac{1}{2}\left( \sigma^2(\Lambda)_1+\sigma^2(\Lambda)_0 \right)}} ,
\label{KC_GEN}
\eeq
\noindent
where $\langle\Lambda\rangle_1$,$\langle\Lambda\rangle_0$ 
are the estimated average LLR over a prescribed $\langle T \rangle$, with and without drive, respectively. We have also denoted with  $\sigma(\Lambda)_1$,$\sigma(\Lambda)_0$ the corresponding estimated standard deviations.
 Index $d_{KC}$ is one possibility among many \cite{Mahalanobis36,KC,Macmillan05} to differentiate between $\mathcal{H}_1$ and  $\mathcal{H}_0$ and to summarize the detection performances in a single value, and therefore often used to quantify SR  \cite{Ward02}.

The use of the $d_{KC}$ index relies on the following considerations.
The performances of a detector are often represented by using the Operating Characteristics (OC) of the detector, e.g. the curve that describes the behavior of $P_m$ as a function of $P_f$.
A first simplification is to compute a particular point of the OC, in which the two error probabilities are equal ($P_f = P_m$), and to  call this value \emph{error probability} $P_e$.
Moreover, under suitable hypothesis and for large sample size $M$,  LLR is asymptotically normally distributed due to central limit theorem (see \cite{Billingsley}). Thus, after simple algebra, we can write:
\beq
P_e=
\displaystyle{\frac{1}{2}\mbox{erfc}
\left(\sqrt{1+\frac{\Delta(\Lambda)^2}{4}} 
\frac{d_{KC}(\Lambda)}{2\sqrt{2}}
\right)},
\label{eq:perrvsdkc}
\eeq
where $\Delta(\Lambda)=2 \mid \sigma(\Lambda)_0-\sigma(\Lambda)_1 \mid/ \mid \sigma(\Lambda)_0+\sigma(\Lambda)_1 \mid$.
Equation (\ref{eq:perrvsdkc}) is a decreasing function of $\Delta$, and therefore neglecting the difference among standard deviations (if it exists) it is possible to retrieve an upper bound of $P_e$ that is only function of $d_{KC}$, i.e.
\beq
P_e\le 
\displaystyle{\frac{1}{2}\mbox{erfc}
\left( 
\frac{d_{KC}(\Lambda)}{2\sqrt{2}}
\right)}.
\label{eq:inequa}
\eeq
The inequality (\ref{eq:inequa}) underlines the heuristic character of the KC index as an indicator of the detector performance.
We also note that $d_{KC}(\Lambda)$ can be derived from the KC index computed on the single $\mathcal{L}_k$, $k \in [0,M-1]$, as 
$d_{KC}(\Lambda) = \sqrt{M} d_{KC}(\mathcal{L})$.

To summarize: the presence of a coherent drive can be detected collecting a suitable numbers of observations, and the performances of the detection can be evaluated using the index given by Eq.(\ref{KC_GEN}). 
In the following subsections we compute the LLR statistic defined in Eq.(\ref{eq:loglikeT}) for two significant types of detection strategies based on the escape times and the strobed dynamics.

\subsection{Escape time sequence-based strategy}
\label{sub:escape}
Let us now analyze detection strategies based on escape time sequences \cite{Dari10}.
More specifically, we examine  the {\em running state} of the bistable oscillator, i.e. we retrieve the escape sequence (driven or not by an external {\hg deterministic signal}) letting the system to freely evolve. 
{\hg A single escape time $\tau_i$ is, loosely speaking,  the time to pass from a basin to the other  (see the Appendix for the details).}

In the described process many time scales occur   \cite{Anishchenko07}. The particle initial position evolves and relaxes in the proximity of the stable points $x=\pm 1$ on a time scale that is governed by the inverse of the friction $\simeq 1/\gamma$. A second time scale is the principal escape rate, as it describes the transition from one side of the barrier to the other.
Finally, a third time scale, the response relaxation time, corresponds to the time necessary to achieve statistical equilibrium, that we refer to as running state.
The running state corresponds to the system free evolution, when the influence of the initial conditions is negligible.
In the following of the paper we consider the sequence of escape times from a basin to the other in statistical equilibrium. 
Details of the numerical procedure to retrieve such sequence taking into account the other time scales is discussed in the Appendix.
The procedure generates an ET sequence $\{\tau_i\}_{i=1}^{\infty}$ that is the starting point of our analysis.

Other choices are possible \cite{Filatrella10}, for instance the system can be prepared in an initial (e.g. $x=-1$) state with a known signal phase $\varphi_0$, and the escape is defined as the shortest time to reach the separatrix. 
Once the separatrix is passed, to prepare the prescribed initial state requires an additional time to set the initial condition at each passage with the required signal phase. 
The additional time should be included in a careful analysis of the performances for signal detection. Moreover, the restart process introduces a further complication in the experimental setup. 
Thus, for sake of simplicity, we only consider the free running dynamics.
 
With the free running procedure one can approximate the probability density of the escape times, $\{\tau_i\}_{i=1}^{\infty}$. 
The distribution depends on the parameters of the {\hg deterministic signal and of the }noise \cite{notePDF}, and it can be computed for each of the two hypothesis $\mathcal{H}_0$ and $\mathcal{H}_1$ to estimate $\widehat{f}_{\tau}(\cdot|\mathcal{H}_0)$ and $\widehat{f}_{\tau}(\cdot|\mathcal{H}_1)$ using a non-parametric statistical technique such as the \emph{Kernel Density Estimation} (KDE) \cite{silverman}, that employs a large number of training samples $\tau_i$ ($> 10^6$ trials) to have stable results (see also \cite{Addesso12}).
A finite sequence $\underline{\tau} = \left\{\tau_k\right\}_{k=0}^{M-1}$ of $M$ retrieved ETs (that plays the role of the observations 
$\underline{y}$) is employed to decide if a coherent  drive is embedded in the perturbation [equationwise, to decide if $\alpha \ne 0$ in Eq.(\ref{signal})]. 
The decision statistic results in the following test:
\beq
\widehat{\Lambda}(\underline{\tau}) = \displaystyle{\sum_{k = 0}^{M-1}\log \left[ \frac{\widehat{f}_{\tau}(\tau_k|\mathcal{H}_1)}{\widehat{f}_{\tau}(\tau_k|\mathcal{H}_0)} \right]}
\test
\zeta.
\label{eq:loglikeEscape}
\eeq

\noindent Thus, it is possible compute the KC index for the ETs: 
\beq
d^e_{KC}  = d_{KC}(\widehat{\Lambda}).
\label{KC}
\eeq
The choice of $M$ is related to the mean escape time without drive $\langle \tau \rangle_0$ and to corresponding mean observation time $\langle T \rangle_0$ by the approximated relation
\beq
\langle T \rangle _0 = M \langle \tau \rangle _0.
\label{eq:CoheTime}
\eeq


\begin{figure*}
\centerline{\includegraphics [keepaspectratio,width=14cm]{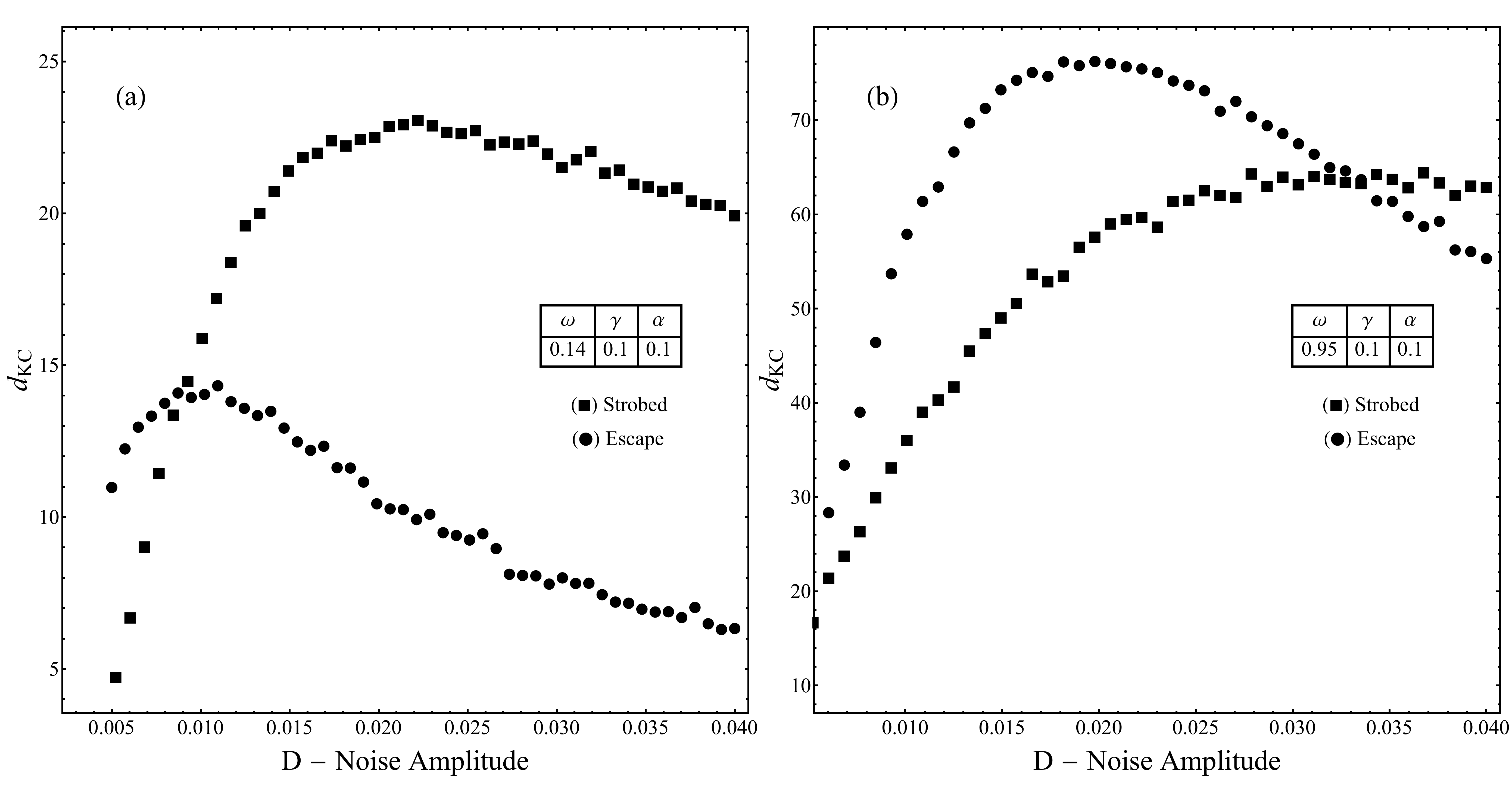}}
\caption{Performances of the detection for the second order system, as measured by the KC index $d_{KC}$, for the ET and SD strategies at the SR frequency (a) and geometric resonance (b). SR occurs at $D\simeq 0.025$, as per Eq.(\ref{eq:SR}), The behavior in (a) is the analogous of SR for first order systems, Fig. \ref{defVSstrobALL}.
The best performances are obtained from the strobing technique in (a), and from the ET in (b).
}
\label{icastica}
\end{figure*}

\begin{figure}
\centerline{\includegraphics [keepaspectratio,width=8cm]{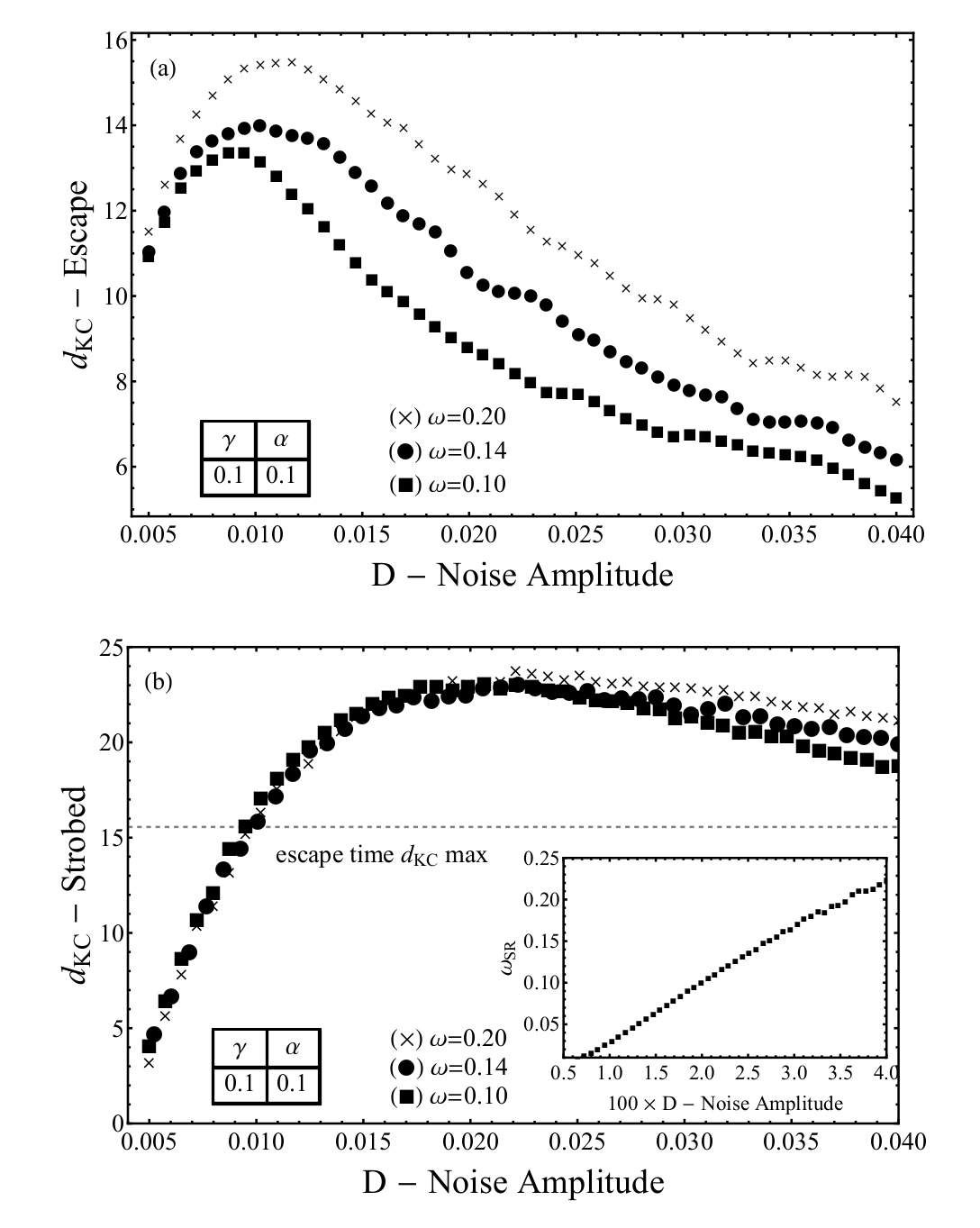}}
\caption{Performances of the ET (a)  and SD (b) strategies of the underdamped system at low external frequencies $\omega$ given by  Eq.(\ref{eq:SR}). 
The position of SR frequency as a function of the noise intensity is reported in the inset. The performances of the SD in (b) are weakly dependent upon the drive frequency, and therefore it is not possible to clearly identify the behavior of the peak.
}
\label{graf2}
\end{figure}

\begin{figure}
\centerline{\includegraphics [keepaspectratio,width=8cm]{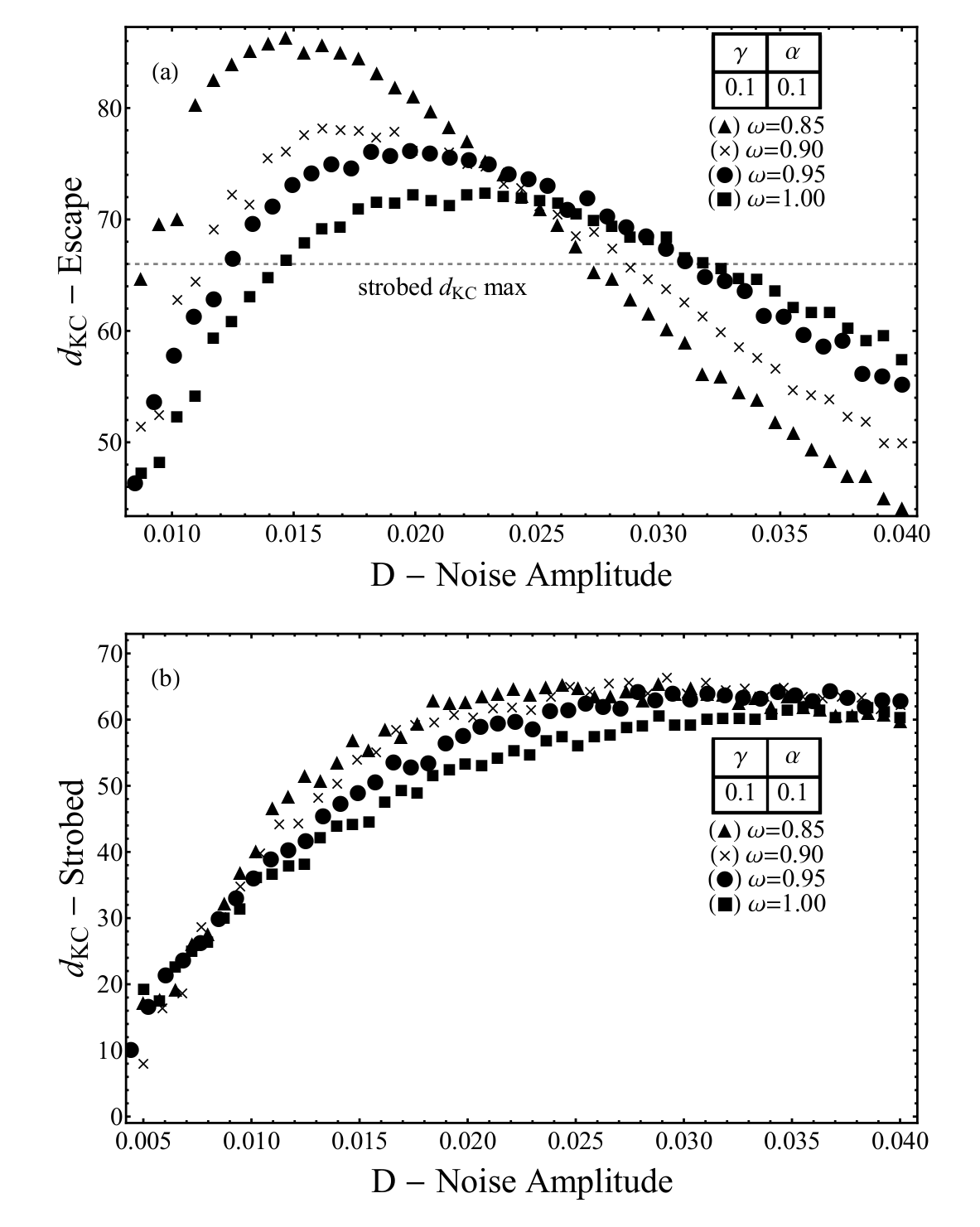}}
\caption{Detailed examination of  $d_{KC}$ for the ET (a) and SD (b) strategies at the geometric resonance for the underdamped system as a function of the {\hg deterministic signal } frequency.  
The dashed line in (a) denotes the asymptotic performances of the strobed technique, as displayed in (b). In (a) the escape strategy performances improve with a suitable choice of the frequency of the {\hg deterministic signal}. The best performances are achieved at the resonance of the system.The performances of the SD detection strategy saturate when noise increases (b).
}
\label{icasticanew}
\end{figure}

\subsection{Strobed Detection Strategy}
\label{sub:strobed}
Strobing amounts to analyze the solution of either Eq. (\ref{eq:bistafirst}) or Eq. (\ref{eq:bistasecond}) at constant time intervals $t=t_k=k \Delta t$, i.e. to construct a sequence $x(t_k)$, to decide if a drive is present. 
The analysis of a strategy based on sampling at constant time intervals leads to the following asymptotic result: for very short strobing time, $\Delta t << 2\pi/\omega$,  the detection amounts to the scalar product (or the Fourier analysis, equivalent to the matched filter), i.e. it is equivalent to the optimal solution \cite{Galdi98}. It is therefore interesting to investigate strobing at finite time intervals. 
It is also known from the theory of SR \cite{Gammaitoni98} that for $\Delta t = \pi/\omega$ (sampling at half the {\hg deterministic signal} period) there exists a suitable phase $\psi$ where the probability distribution of the sequence $x(t_k)$ peaks at $x=1$ for even $k$, and at $x= - 1$ for odd $k$. Formally, computing $N = \langle T \rangle/T_s$, where (as usual) $\langle T \rangle$ is the mean observation time and  $T_s=2\pi/\omega$, it is possible to define $2N$ sampling times: 
\beq
t_k =\frac{1}{\omega}
\left(
\frac{2k+1}{2}\pi-\psi
\right), ~~~~~k=0,1,2,..., 2N-1
\label{eq:strob_tim}
\eeq
\noindent to obtain the sequence: 
\beq
x_k=(-1)^k x(t_k).
\label{eq:sequence_strobed}
\eeq

In a perfectly synchronized state the system crosses the separatrix each half period; if this is the case, the system exactly jumps from $-1$ to $1$, and back from to $1$ to $-1$. Thus the sequence (\ref{eq:sequence_strobed}) only contains positive elements. On the contrary, for a purely random sequence  that casually moves in the phase space, the sequence (\ref{eq:sequence_strobed}) contains as many positive as negative symbols. Put in another way, using the Heaviside $\Theta$ function it is possible to obtain the
observations 
\beq
s_k = \Theta(x_k).
\eeq
These observations are a Bernoulli random variable with parameter $P_+=\mbox{Prob}(s_k = 1)$.
Thus, the LRT decision statistics is given by counting the number of plus signs, i.e.
\beq
N_+=\sum_{k=0}^{2N-1} s_k,
\label{eq:known_phi}
\eeq
where the number of observation is $M = 2N$.
As usual, the statistics $N_+$ can be compared to a threshold $\Gamma$ to decide if a signal is present, in that one expects $N_+= N$ for a purely random signal, and $N_+= 2N$ for a sinusoidal drive without noise. The above statistics is referred to as strobed sign-counting  stochastic resonance detector (SSC-SRD) \cite{Galdi98}.

The SSC-SRD performance is described by the false-alarm and false dismissal probabilities \cite{Galdi98}:
\beq
P_f=\mbox{Prob}
\left\{
N_+ > \Gamma | \mathcal{H}_0
\right\}=
I_{1/2}(\Gamma+1,2N-\Gamma),
\eeq
$${} $$
\beq
P_m=\mbox{Prob}
\left\{
N_+\leq \Gamma | \mathcal{H}_1
\right\}=
1-I_{P_+}(\Gamma+1,2N-\Gamma),
\label{eq:ROCs}
\eeq
where $I_p(x,y)$ is the regularized incomplete Beta function of order $p$. 
The related OCs are typically worse by $\approx 3dB$ than the corresponding OCs of the matched filter \cite{Galdi98}.

Also in this case, for high values of $2N$ it is possible to approximate the binomial distribution via a normal one (by using the well-known de Moivre-Laplace theorem \cite{FellerV1}) and to introduce a suitable KC index for the strobed statistics, i.e. 
\beq
d^s_{KC} = d_{KC}(N_+).
\label{KC_S}
\eeq
It is noticeable that in this case there is a simple expression for $d^s_{KC}$, i.e.
\beq
d^s_{KC} = \sqrt{2N} \left[\frac{2 P_+ -1}{\sqrt{\frac{1}{2}+2 P_+ (1-P_+)}}\right].
\label{eq:dkc_strobed}
\eeq

\begin{figure}
\centerline{\includegraphics [keepaspectratio,width=8cm]{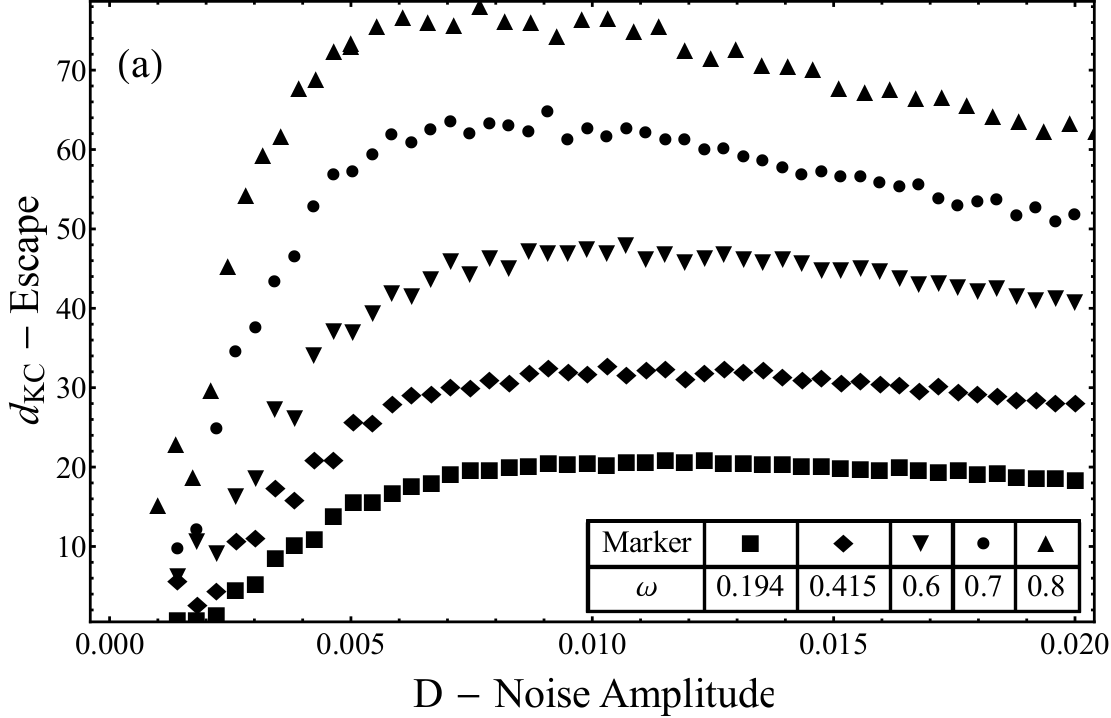}}
\centerline{   \includegraphics [keepaspectratio,width=8cm]{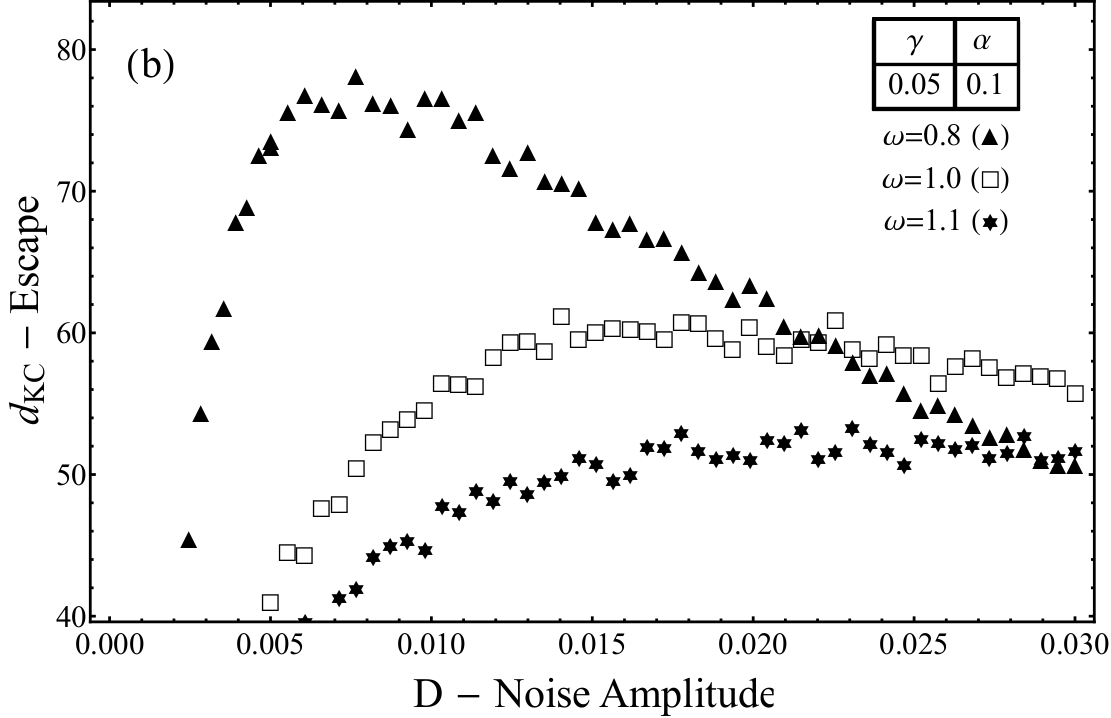}}
 \caption{Underdamped escape times strategy  for various external frequencies $\omega$. 
 The best performances are observed at $\omega=0.8$ in (a) \cite{Addesso12b}, in fact the peak performances increase in (a) ($\omega \le 0.8$), and decrease in (b) ($\omega \ge 0.8$).
This is in contrast with the monotonic behavior of the relation (\ref{eq:SR}): the temperature at which the peak occurs for different frequencies first increases (a) and then decreases (b).   
This clearly defines a standard geometric resonance with a favorite escape time.
The first order analogue, Fig.\ref{defVSstrobALL}, is monotone in the external frequency $\omega$. 
}
\label{fig7}
\end{figure}

\begin{figure}
\centerline{\includegraphics [keepaspectratio,width=8cm]{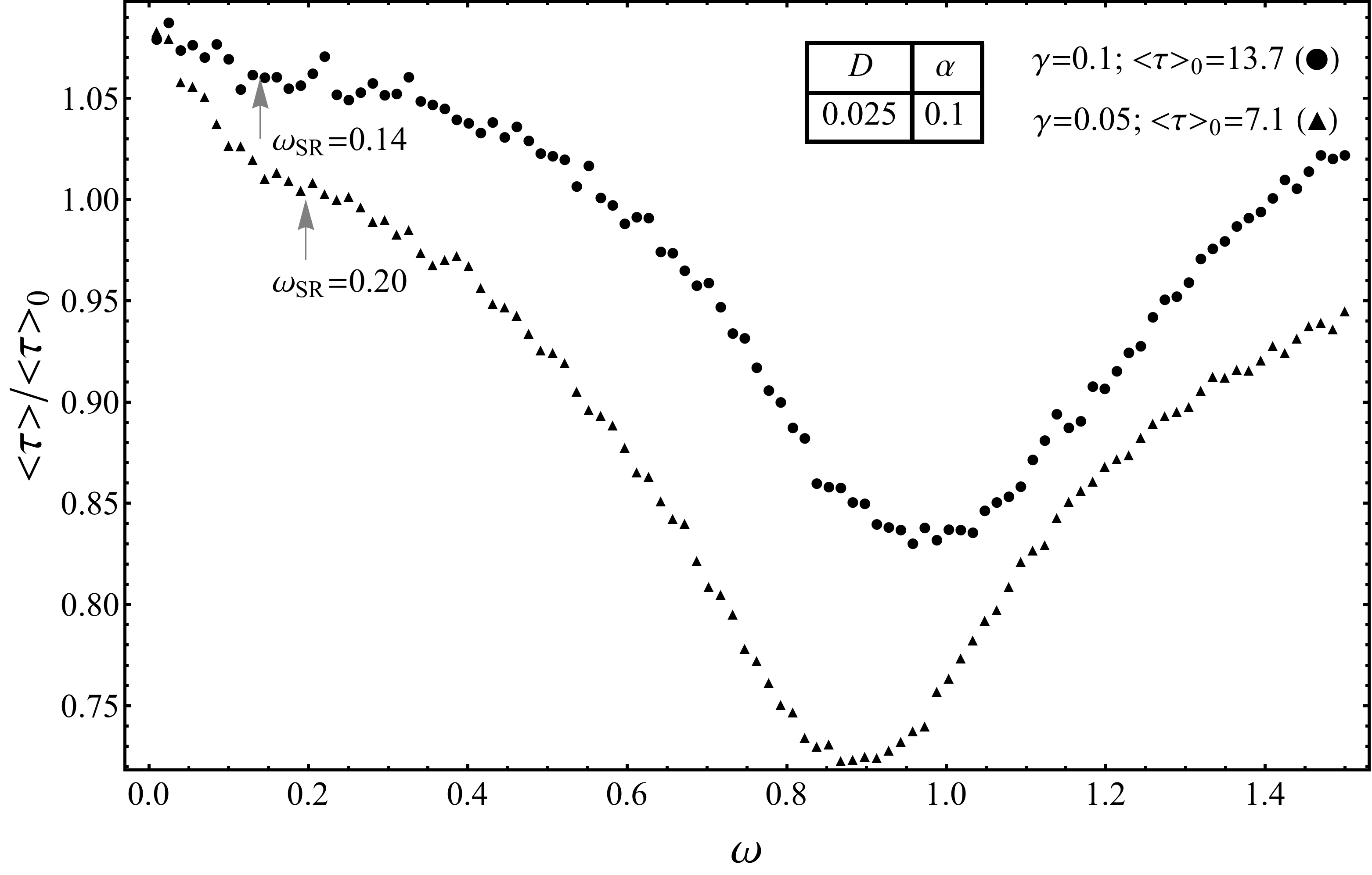}}
\centerline{\vspace{-2.5cm}}
\centerline{(a) \hspace{6cm}}
\vspace{2cm}
\centerline{ 
  \includegraphics [keepaspectratio,width=8cm]{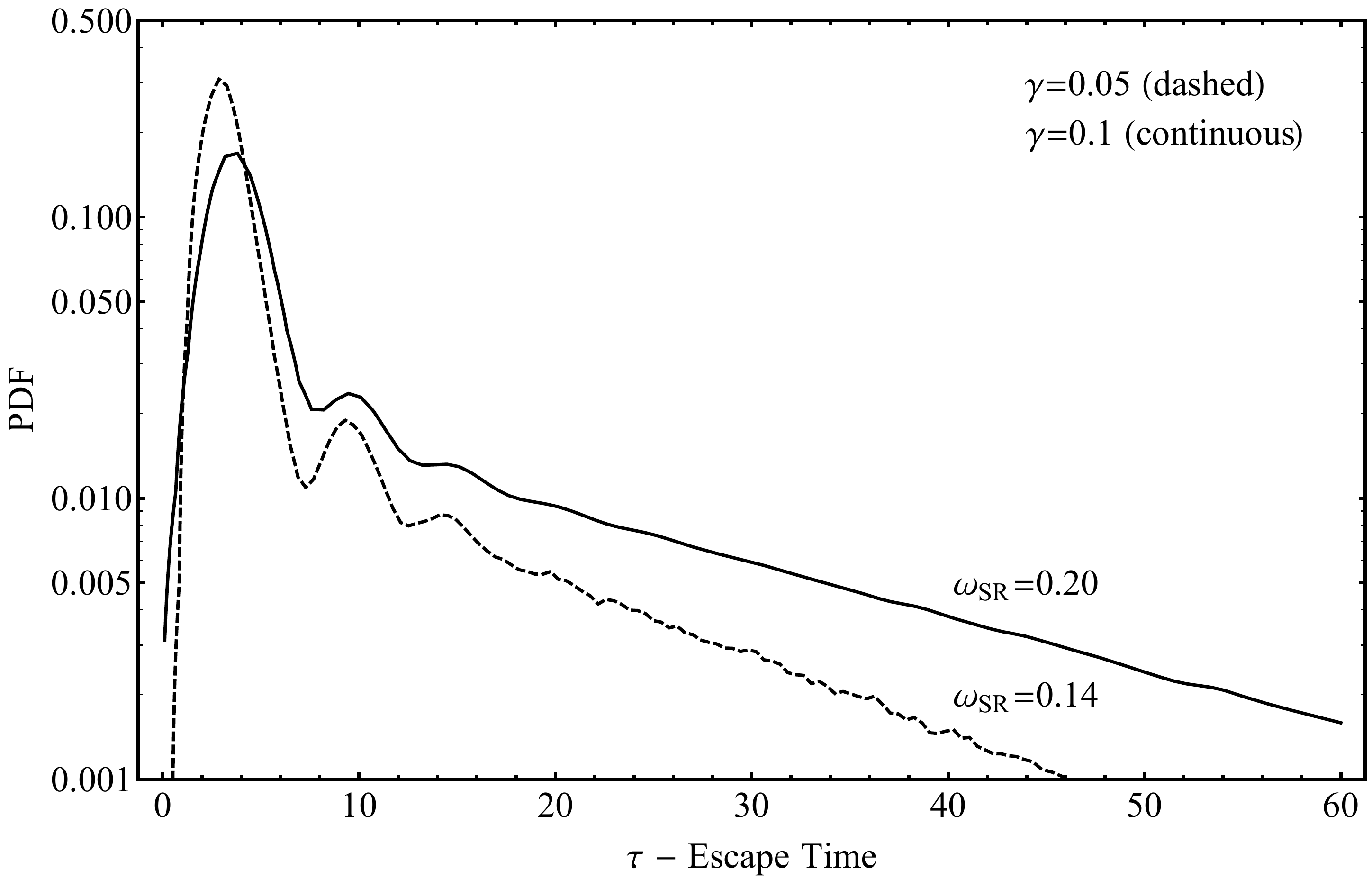}
}
\centerline{\vspace{-2.5cm}}
\centerline{(b) \hspace{6cm}}
\vspace{2cm}
\centerline{ 
  \includegraphics [keepaspectratio,width=8cm]{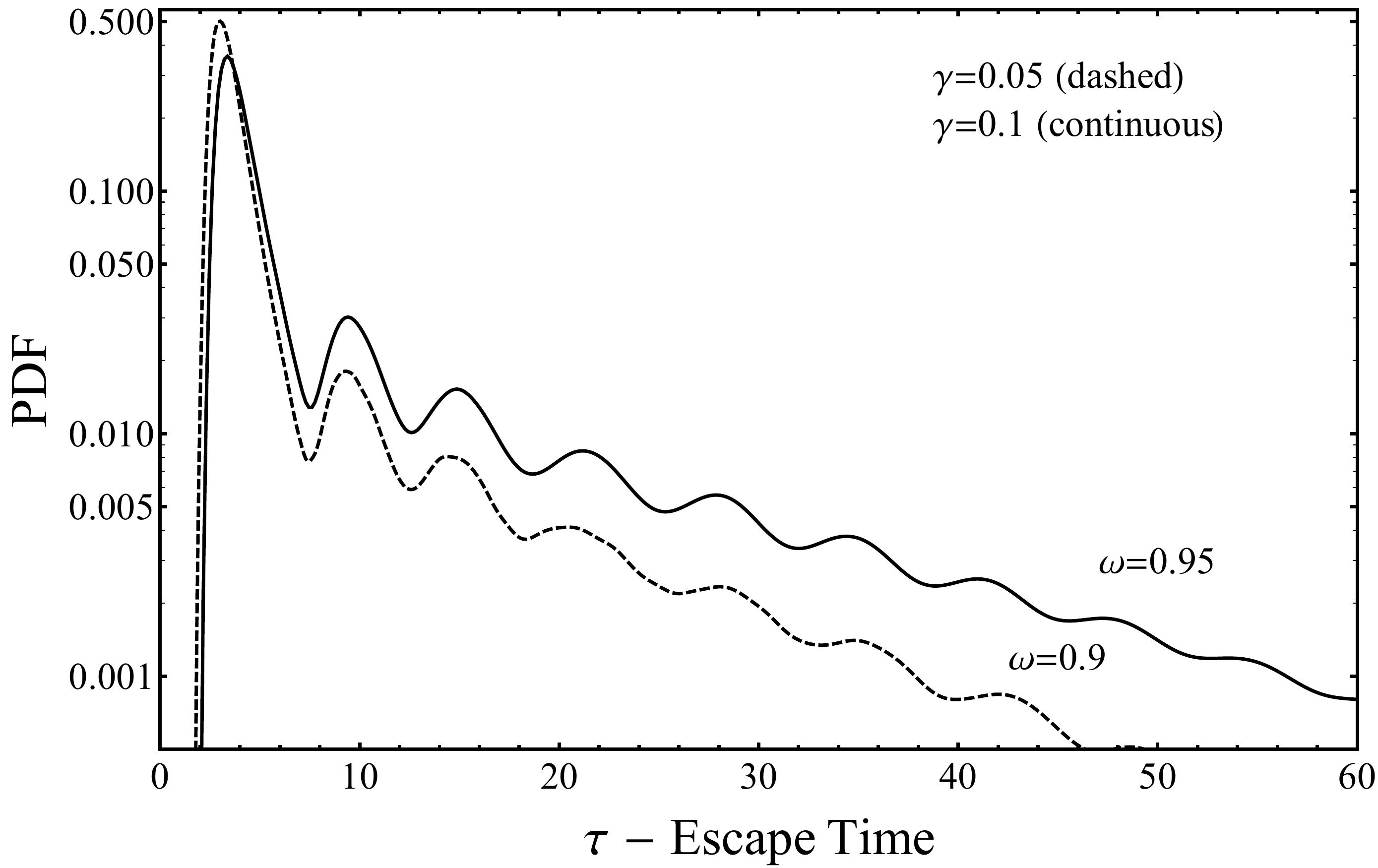}
}
\centerline{\vspace{-2cm}}
\centerline{(c) \hspace{6cm}}
\vspace{1cm}

\caption{(a) The  frequency dependence of the normalized escape time of the underdamped system for two values of the dissipation parameter ($\langle \tau \rangle _0$ is the average escape time in {\hg the presence of a purely random  signal}). 
The dip occurs at the geometric resonance. 
At the SR frequency (estimated by the asymptotic  behavior of the ET distribution) we underline a weak change in concavity of the curve, indicated by the arrows.  
 (b) PDFs of the escape time,  at the frequency $\omega = \omega_{SR}=\pi/\langle\tau\rangle_0$.
(c) PDFs of the ETs at the geometric resonance frequency.
}
\label{avetau}
\end{figure}

Finally, it is possible to design a detection strategy when the phase $\psi$ is unknown, that relies on Generalized LRT (GLRT)  approach, in which a filter bank jointly performs the estimation and detection tuning the initial phase $\psi^{h}=2\pi h/H$, where $h = [0,H-1]$.
Thus, it is possible to collect a vector of likelihood ratios and the final decision statistics is obtained comparing the maximum value of the detectors with a threshold $\zeta$:
\beq
\max_{h \in [0,H-1]}
\sum_{k=0}^{2N-1}
s_k(h)
 \test
 \zeta,
\label{eq:pick_max}
\eeq
where the observations are
\beq
s_k(h) = \Theta \left[
(-1)^k x
\left(
t_k+\frac{hT_s}{H}
\right)
\right]
\eeq
(again $T_s=2\pi/\omega$).
It has been shown that the resulting  unknown-initial-phase  detector for  $H > 10$
has nearly {\em the same} performance as  (\ref{eq:known_phi}), which applies to the coherent (known initial phase) case \cite{Galdi98}, and it is comparable to the performances of the {\em noncoherent} correlator  (standard optimum   benchmark detector for signals  with unknown initial phase).
On the other hand, the SSC-SRD is computationally {\em extremely cheap}, in that it only requires binary and/or integer arithmetics.

\section{Detection performances}
\label{sec:performances}
From this Section we begin to collect a systematic analysis of the performances of the detection realized through the analysis of the ET and the strobed sequence, see Sect. \ref{sec:detection}. 
The numerical simulations have been performed with Euler algorithm taking into account the correction proposed in Ref. \cite{Mannella00}. 
The convergence of the results has been checked both decreasing the integration step size and increasing the simulation time. 
Typical  results have been obtained with very long simulations (number of trials for PDF estimation is $\sim 10^6$).
The observation time  (after a transient that has been discarded) for signal detection has been set to $\langle T \rangle \simeq 4 \cdot 10^5$.
The running state analyzed is independent of the initial conditions (initial phase $\varphi_0$ and position), and therefore there is no need to replicate the system.  

\subsection{First order (overdamped) system}
\label{sub:overdamped}

For the overdamped system governed by Eq.(\ref{eq:bistafirst}) the KC indexes $d^e_{KC}$ and $d^s_{KC}$, given by Eqs. (\ref{KC}) and (\ref{KC_S}), show a remarkably different behavior for the two data acquisition strategies, see Fig. \ref{defVSstrobALL}. 
It is important to notice that the underlying trajectories to be analyzed are the same; the two detection strategies are just two different ways to reduce the data, either to a sequence of passage times (ET)  or to sample the position at constant time intervals (SD). 

From Fig. \ref{defVSstrobALL} it is also clear that also the best performances of the detection strategy based on the ETs are below the performances of the strobing  in the considered noise interval (we measure the quality of the strategies through the $d_{KC}$ peak height, and therefore  secondary peaks are less interesting).
The detection performances of the SD deteriorate when the frequency increases, while ET performances remain nearly unchanged.
As shown in Fig. \ref{indicaSRdiversi}a, this is consistent with an analysis of the displacement amplitude $\bar{x}$ implicitly defined by \cite{Gammaitoni98}:
\beq
\langle x(t) \rangle = \bar{x} \sin \left( \omega t + \bar{\Phi} \right).
\label{amplitude}
\eeq

 \noindent 
The quantity $\bar{x}$ is approximately given by the Equation:
\beq
\bar{x} = \frac{2 \alpha \langle x^2 \rangle _0 }{ D\sqrt{4+\omega^2 \langle \tau \rangle_0^2 }}
\label{displacement}
\eeq
where $\langle x^2 \rangle_0 $ is the position variance without sinusoidal drive ($\alpha = 0$).
Figure \ref{indicaSRdiversi}b sheds light on the physical origin the performances of the detection of Fig. \ref{defVSstrobALL}b: at the noise level  $D\simeq 0.2$ the strobed KC index $d^s_{KC}$ exhibits a peak in Fig. \ref{defVSstrobALL}b and in the same position the amplitude $\bar{x}$, Eq. (\ref{displacement}), also exhibits a peak. 
A peculiar behavior for the same frequency is also observed for the distribution width $\rho$, that quantifies the degree of synchronization with the input {\hg periodic drive} \cite{Wellens04}:
\beq
\rho = \sqrt{\frac{\langle \tau^2 \rangle }{\langle \tau \rangle ^2}-1}
\label{rho}
\eeq
(here $\langle \tau^2\rangle $ and $\langle \tau \rangle $ are the ETs moments with the applied sinusoidal drive).
The index (\ref{rho})  measures deviations from the exponential case ($\rho = 1$) and it is often taken as a signature of SR. However, the distortion of the escape time distributions of Fig. \ref{indicaSRdiversi}b are not fateful for signal detection.
The analysis of the full distribution of the ETs embedded in Eq. (\ref{KC}), not just of some parameters as in Eqs. (\ref{displacement}) and (\ref{rho}), does not show a nonlinear behavior, see Fig. \ref{defVSstrobALL}b.
Thus a physical stochastic resonance, such as a peak in the displacement, Fig. \ref{indicaSRdiversi}a, or a change in the distribution of the escape times, Fig. \ref{indicaSRdiversi}b, does not necessarily imply an improvement of the ETs detection properties given by Eq. (\ref{eq:locloglik}).

We conclude that in the prototypical overdamped bistable system, if one wants to exploit  SR for signal detection, the details of the detection strategy are essential. 
The analysis of the ETs cannot be improved by the increase of noise, while the analysis of the strobed dynamics does show an improvement at the SR frequency.

\subsection{Second order (underdamped) system}
\label{sub:underdamped}
We continue our analysis of the performances of the detection, and now discuss the case of an underdamped system. 
Two regions are a priori interesting: the SR frequency where the relation Eq.(\ref{eq:SR}) holds ( $\omega \ll 1$ for these parameters), and the geometric resonance, or the oscillation frequency at the minimum of the potential ($\omega \simeq 1$ in these normalized units). 
{ \hg 
The two frequencies are related to interwell oscillations  (here corresponding to $\omega \ll 1$) and intrawell regimes (here corresponding to $\omega \simeq 1$). 
}
Typical results are collected in Fig. \ref{icastica}a,b, that is the main result of this paper. Figure \ref{icastica} indicates that there are some striking differences with the analogous behavior for the overdamped case of Fig. \ref{defVSstrobALL}. 
First, at the lower frequency corresponding to the SR resonance,  the $d_{KC}$ peak of strobing better performs compared to the  peak of  ET, see Fig. \ref{icastica}a. 
At higher frequencies corresponding to the geometric resonance, a peak appears that it is only present in the underdamped system, see Fig. \ref{icastica}b.
This peak indicates that the ETs better perform respect to the strobed strategy: at the geometric resonance, Fig. \ref{icastica}b, the role of the best detection strategy is exchanged respect to the SR frequency of Fig. \ref{icastica}a.
Thus depending on the nature of the resonance (stochastic or geometric) the detection strategy that exhibits the best performances is changed. 
We ascribe the behavior to the different nature of the two peaks:
the SR low frequency peak is a synchronization phenomenon between noise and deterministic drive that does not accumulate energy in the oscillations. 
In contrast, the geometric resonance is a {\it standard} resonance due to energy storage of the oscillations in the well (as such, it cannot be observed in the first order system). 
Thus, the two strategies, SD and ET, are best suited for the stochastic and geometric resonance, respectively. 
The two peaks of the strategies are found at different noise level (Fig. \ref{icastica}), though all parameters are the same. This is possible inasmuch  the index $d_{KC}$, Eq.(10), exploits two data sets: one in presence of noise alone ($\mathcal{H}_0$) and one in the presence of the signal $S(t)$ ($\mathcal{H}_1$). A maximum occurs when a strategy enhances the difference, and hence the maximum depends upon the details of the employed strategy.
In contrast, if one considers SR measures such as $\rho$ or $\bar{x}$,  Eqs. (26) or (28), only based upon data sequences containing the signal ($\mathcal{H}_1$), noise most affects the sequences at a specific noise level, likely to be the same for any indicator (Fig. 3b).  

The analysis of Fig. \ref{icastica} of the detection features for different frequencies of the external drive are further explored in Figs. \ref{graf2},\ref{icasticanew}, and \ref{fig7}.
In the neighborhood of  $\omega_{SR}$, Fig. \ref{graf2}, the performances behavior as a function of the noise identifies a peak for the escape based detection, that we compare with the frequency  $\omega_{SR}$ predicted in Eq.(\ref{eq:SR}).
The noise intensity at the peak of Fig. \ref{graf2}a, collected in the inset of Fig. \ref{graf2}b, moves to higher noise level when the frequency is increased, as predicted by Eq.(\ref{eq:SR}) that captures the order of magnitude of the position of the best performances.
We thus confirm that SR, the synchronization of the external drive of noise-induced escapes, seems to be the physical origin of the peak; however SR theory cannot be used for a detailed prediction of the best noise input.
We remind that in the SR region the peak of the ETs in Fig. \ref{icastica}a is not useful for detection, inasmuch the performances are inferior to the strobed performances.

In Fig. \ref{icasticanew} we investigate the physics and the detector performances around the geometric resonance. 
The peak of the detection performances moves to higher noise when the drive frequency increases for the ETs (Fig. \ref{icasticanew}a), and stays at almost the same frequency for the strobed strategy (Fig. \ref{icasticanew}b) --the slight change is due to the weak change of the resonant frequency, that for nonlinear systems is affected by noise and other parameters \cite{Addesso12b}.

To connect the behavior from low to high frequency (i.e., from the SR to the geometric resonances) we plot the  performances of  the ETs based strategy in Figs. \ref{fig7}  as a function of the noise temperature for different frequencies. 
It is noticeable that, at variance with the theoretical behavior of the standard SR Eq.(\ref{amplitude}) and with the performances of the detection of overdamped systems (Figs. \ref{defVSstrobALL}a,b, and c) the dependence upon the external frequency exhibits a maximum around $\omega \simeq 0.8$. 
Also, in contrast with the monotonic behavior of the relation (\ref{eq:SR}), the temperature at which the peak occurs for different frequencies first increases (Fig. \ref{fig7}a) and then decreases (Fig. \ref{fig7}b).  

{\hg We underline that the geometric resonant frequency, at which the index $d_{KC}$ peaks, is sensitive to dissipation $\gamma$, for the response of the system is nonlinear. 
However, the actual peak  frequency stays below unity  \cite{Addesso12}.
}
The frequency dependence is also analyzed through Fig. \ref{avetau}a, that displays the physical average escape time as a function of the {\hg deterministic signal} frequency. 
When the drive frequency is in the neighboring of the geometric resonance at $\omega \simeq 0.8$, the escape times are shortest, for the average $\langle \tau \rangle $ mostly deviates from the average escape time without {\hg deterministic signal} $\langle \tau \rangle _0$. In this condition the performances of the ET strategy are maximized, see Fig. \ref{fig7}b.

The PDFs of Figs. \ref{avetau}b,c display other important physical features that affect detection. At the lower frequency, corresponding to SR {\hg, due to the phenomenon of resonant activation,} the PDF is almost exponential {\hg with a cutoff \cite{mantegna98a}}. 
At the geometric resonance{\hg, due to the phenomenon of dynamical resonant activation}, the average escape changes and the distribution develops large oscillations, see Fig. \ref{avetau}c, around the exponential behavior of the unperturbed ($\alpha = 0$) system. 
The oscillations reveal that the interwell transitions occur at a specific  value of the signal phase (as observed, for instance, with the incoherent acquisition strategy and initial conditions reset in Ref. \cite{Addesso12b}). 
{\hg This is analogous to the oscillations found in experiments on overdamped dynamics of tunnel diodes \cite{Mantegna98b}.

In the case of a weak sinusoidal drive, that is the interesting limit for signal processing, the signal only introduces a small deviation from the unperturbed case instead of a secondary minimum \cite{Valenti14}.
Actually, an inflection point, indicated by an arrow, appears in Fig. \ref{fig7}a.
}

The decision test, based on Eqs. (\ref{eq:loglikeEscape}, \ref{KC}), exploits the oscillations of the PDF to distinguish between the perturbed and the unperturbed PDFs, and it is therefore most effective at the geometric resonance. 

\begin{figure}
\centerline{\includegraphics [keepaspectratio,width=8cm]{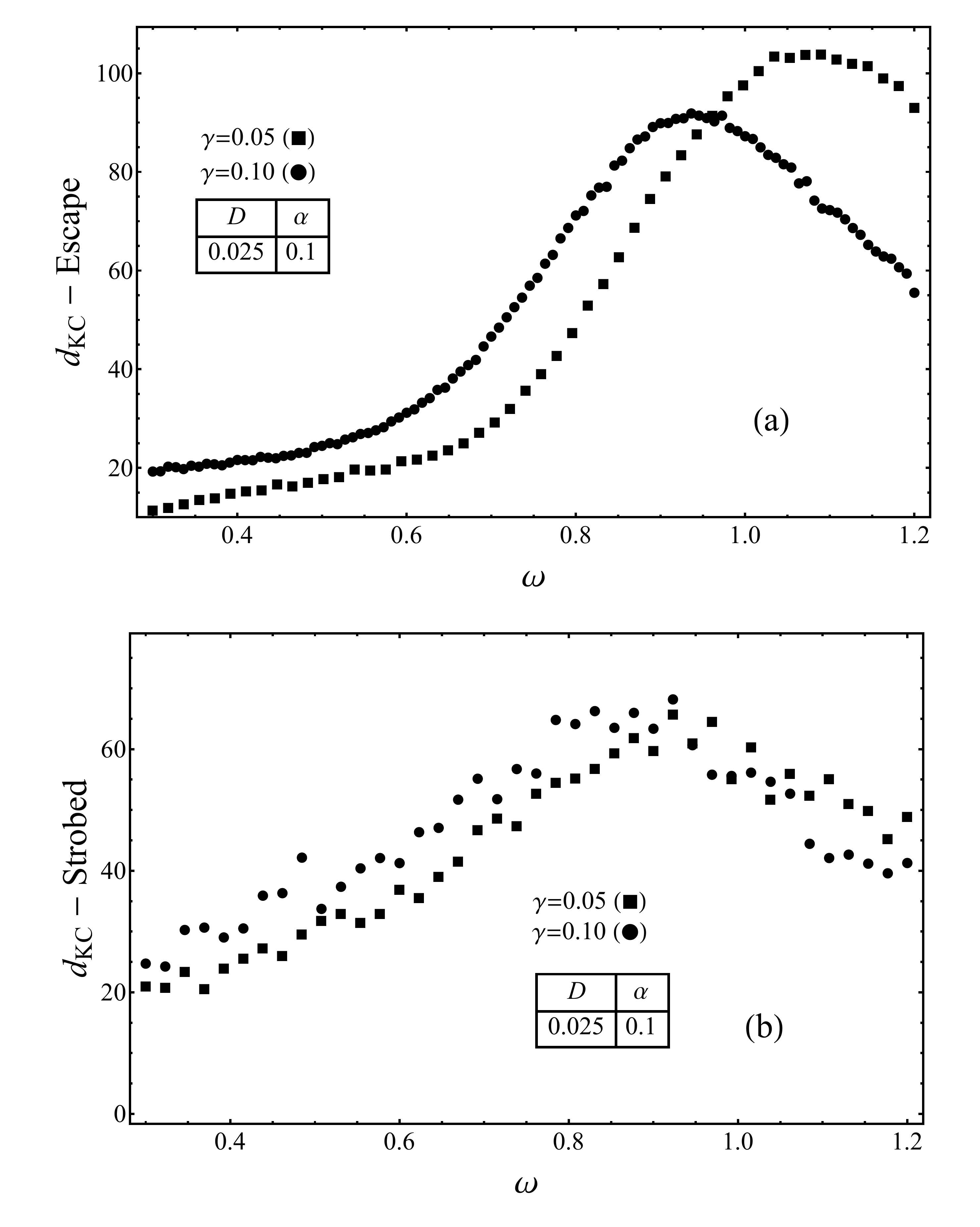}}
\caption{Underdamped system,  (a) escape and (b) strobed signals for two values of the dissipation $\gamma$. 
The escapes peak strongly depends upon the dissipation $\gamma$. 
Instead, the strobed strategy is little affected by dissipation, as expected for a first order type detection.
}
\label{fig9}
\end{figure}

Finally, in Fig. \ref{fig9} we investigate the role of dissipation. Figure \ref{fig9}a  confirms that the detection thorough the escape times is favored at low dissipation -- indeed for overdamped systems it is passed by the strobed strategy. In contrast, the strobed based strategy at the lower frequency is little affected by dissipation, see Fig. \ref{fig9}b.

\section{Conclusions}
\label{sec:conclusions}
Stochastic resonance,  that is an enhancement of the coherent response of a system in spite of an increase of the noise, is viewed through the performances of the detection analysis. 
In this framework  it is known that, adopting optimal detection strategies that exploit the full trajectory (i.e.,the matched filter), the  detection can only improve when noise decreases .
If instead one is forced to some suboptimal analysis on reduced data, as escape times or trajectory discretization (strobing), it has been shown that a trustworthy improvement of the performances could occur even in spite of an increase of the disturbing noise \cite{Kay00,Rousseau05,Ward02}. 

This is the premise to investigate the interaction between stochastic resonance in the signal detection acception and the physical properties of the system. We first construct the likelihood ratio test schemes that can be applied to the escape times and the strobed discretization. 
(We note that escapes and strobing are, to the best of our knowledge, the only discrete data sample strategies considered in the literature.)
The measured performances show that: 
\begin{itemize}
\item[1)] For overdamped and underdamped systems, around the stochastic resonance frequency of Eq. (\ref{eq:SR}), strobing overperforms the escape times based strategy 
\item[2)]  For underdamped systems, at the geometric resonance frequency, the escape times based strategy overperforms strobing; 
\end{itemize}

The results reported in 1) are physically consistent with the synchronization of the periodic drive and the  noise assisted leaps over the energy barrier. 
When the drive period and the noise induced jumps are comparable, the escapes distribution just remain exponential, see Fig. \ref{avetau}b. 
The escape analysis is therefore less effective, as shown in Figs. \ref{defVSstrobALL},\ref{icastica}a. 
Moreover, this type of synchronization can occur for both overdamped and underdamped systems, as shown in Figs. \ref{defVSstrobALL},\ref{icastica}a, confirming that stochastic resonance is inherently a first order phenomenon, also in inertial systems.

The behavior at the geometric resonance, point 2), has also a physical origin: it corresponds to the possibility to accumulate energy in an eigenmode of the potential. 
When this condition occurs, i.e. when the periodic drive frequency matches the natural frequency of the potential, the distribution of the passage times exhibits clear oscillations at the drive period, see Fig. \ref{avetau}c, and the analysis of the escapes  is most effective, see Fig. \ref{icastica}b. 
We remark that the abovementioned detection opportunity only arises in underdamped systems that can accumulate energy:
The performances of the escapes based strategy improve when dissipation decreases, see Fig. \ref{fig9}a, while the strobed based detection does not, see Fig. \ref{fig9}b. 
Shortly, the physical properties (dissipation, characteristic frequency) and the detection strategy are not independent, but deeply intertwined.

{\hg One can conjecture that the prescriptions we have found for the analysis of the reduced data can be extended to other systems as washboard potential \cite{Marchesoni97} or piecewise linear barriers \cite{Agudov10}, that might prove convenient for analytic treatment. }
As a last remark, let us recall that the above results are based on a prototypical bistable quartic potential. 
Results on Josephson junctions \cite{Addesso12,Addesso12b} and Fabry-Perot pendular interferometers \cite{Addesso13} are consistent  with the above interpretation.

\section*{Acknowledgement}
We wish to thank INFN ({\it Napoli, Gruppo collegato di Salerno}) for partial support and from PON Ricerca e Competitivit\`a 2007-2013 under grant agreement PON NAFASSY, PONa3\_00007.

\begin{figure}
\centerline{\includegraphics[scale=0.4]{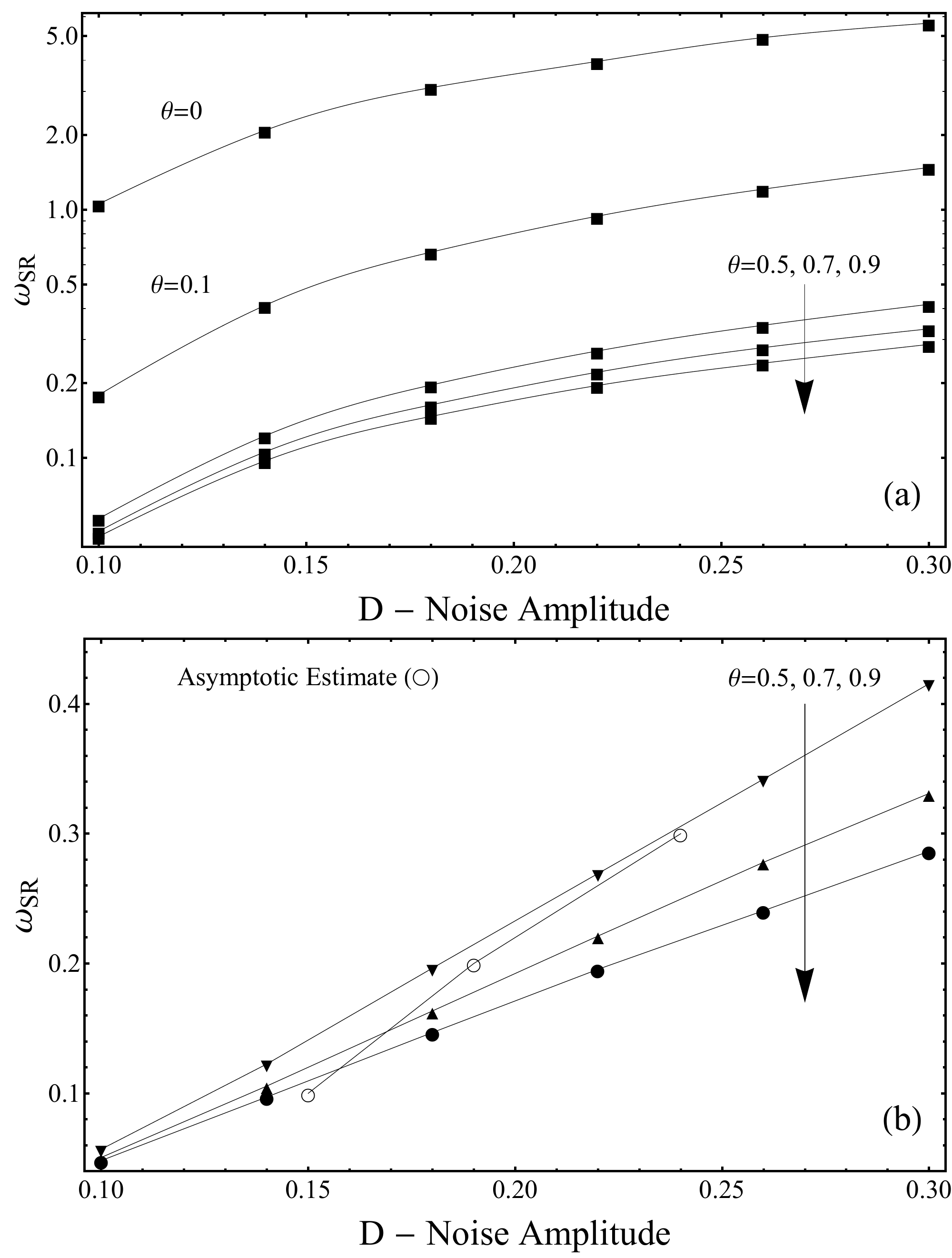}}
\caption{Threshold setting procedure for the parameter $\theta$ of Fig. \ref{fig:potential}.
(a) estimated stochastic resonance frequency [Eq. (\ref{eq:SR})] as a function of the threshold $\theta$.
(b) closeup of$o.5 \le \theta \le 0.9$ (filled symbols) compared with the estimate of $\omega_{SR}$ from the asymptotic exponential behaviuour of the simulated ETs probability distribution (open circles).
}
\label{stimaSR}
\end{figure}

\section*{Appendix - Numerical procedure to retrieve the Escape Time sequence}

It is well known that to determine the passage across a boundary with simulations of  stochastic differential equations is problematic \cite{Mannella00}. 
In this Appendix we sketch the method employed to retrieve the ETs  avoiding spurious fluctuations across the separatrix around the unstable point of the potential (\ref{eq:bistapot}), $x=0$. 
In fact, subject to fluctuations, the representative coordinate $x(t)$ might cross several times the separatrix, fast oscillating around the maximum of the potential before to slip down towards a minimum. 
To avoid such deceitful passages that alters the high frequency part of the escape times spectrum, we use an effective threshold value $\pm \theta$,  see Fig. \ref{fig:potential}. 
If the particle is initially in the left hand side of the potential, ($x<0$), we only count a passage across the threshold $+\theta$, beyond the separatrix in the descending part of the potential \cite{Yamapi14}. 
If instead the system is initially in the right hand side of the potential ($x>0$), an escape is defined as the passage across $-\theta$.

{\hg 
The iteration of the previous rules generates an ET sequence $\{\tau_i\}_{i=1}^{\infty}$ that is the starting point to estimate the escape time distribution at statistical equilibrium, i.e. in the running state.
The choice of the threshold to decide if the passage to the other basin has occurred affects the statistical properties of the escape times sequence.
}
{\hg 
The data are recorded when the system has reached the running state; the procedure amounts to define an escape time as the time to reach the threshold $\theta$  ($-\theta$) starting from the position $x= - \theta$ ($x=\theta$ in the basin $x<0$ ($x>0$).

For the first order system this is analogous to the first passage time from $-\theta$ to $\theta$ (or $\theta$ to $-\theta$ for the reverse passage).
For second order systems the analogy with first passage times is incomplete, as the initial condition on the velocity is not reset.
In fact, for a rigorous mathematical definition of the ETs it is important the {\it restart process}, i.e. to specify the initial conditions after each passage (see Ref. \cite{Fiasconaro03} for the role of initial conditions). 
}

To determine a suitable value for the parameter $\theta$, we observe that the PDF for long escape times values is asymptotically  exponential (as expected in the absence of an applied external drive, see Fig. \ref{avetau}b,c), while for low escape time it deviates from the exponential distribution.
In accordance with these observations, we chose the appropriated $\theta$ value imposing that $\omega_{SR}$, numerically obtained  from Eq. (\ref{eq:SR}),  agrees with the asymptotic exponential decay \cite{Addesso15}.

In Fig. \ref{stimaSR} a typical estimate of stochastic resonance frequency, computed for the second order system (\ref{eq:bistasecond}),  is displayed.
The Figure shows that the frequency of the SR (at a fixed noise level $D$) depends upon the choice of the threshold $\theta$ to discriminate a passage from a basin to the other. 
In particular Fig.\ref{stimaSR}a shows that for $\theta >0.5$ a stable $\omega_{SR}$
estimated can be achieved. 
In the close up showed in Fig.\ref{stimaSR}b the estimate is compared with the asymptotic exponential value of PDF extracted by the simulated  ETs distribution; we note that a reasonable agreement is found for $\theta \simeq 0.6$, that is the value employed in the simulations.

{\hg 
To guarantee statistical independence from initial conditions, the results are collected after a transient time, typically of $T_{trans} \simeq 1000\langle \tau \rangle$, where $\langle \tau\rangle$ is defined by Eq.(\ref{eq:SR}).
We have checked that such time is longer than the typical response relaxation time of the system.
}

Finally, we observe that results similar to Fig. \ref{stimaSR} can be found for the first order bistable system.

\bibliography{etvsst_arxive}

\begin{thebibliography}{57}
\bibitem[{not(sion)}]{notePDF}
\bibinfo{year}{The actually retrieved PDF also depends on the choice of the
  parameter $\theta$. To reduce the spurious dependence form the threshold a
  good choice may be $\theta \simeq 0.6 $ (see the Appendix for details and
  discussion).}
\end{thebibliography}

  \end{document}